\documentclass{ws-p8-50x6-00}
\usepackage{hyperref}
\usepackage{axodraw}
\usepackage{latexsym}
\input boxedeps
\SetepsfEPSFSpecial %%%%% Uncomment for submission
\HideDisplacementBoxes

\renewcommand{\rightnote}{\textsf{TASI 2000 Lectures on Electroweak 
Theory: FERMILAB--CONF--01/001--T} }

\newcommand{\s}{\hbox{ s}}
\newcommand{\etal}{{\em et al.}}
\newcommand{\ie}{{\em i.e.}}

\newcommand{\gevcc}{\hbox{ GeV}\!/\!c^2}
\newcommand{\gev}{\hbox{ GeV}}
\newcommand{\gevc}{\hbox{ GeV}\!/\!c}

\newcommand{\evcc}{\hbox{ eV}\!/\!c^2}
\newcommand{\mev}{\hbox{ MeV}}
\newcommand{\mevcc}{\hbox{ MeV}\!/\!c^2}
\newcommand{\tev}{\hbox{ TeV}}

\newcommand{\tevcc}{\hbox{ TeV}\!/\!c^2}

\newcommand{\fm}{\hbox{ fm}}
\newcommand{\cm}{\hbox{ cm}}
\newcommand{\mm}{\hbox{ mm}}
\newcommand{\nb}{\hbox{ nb}}

\newcommand{\fb}{\hbox{ fb}}

\newcommand{\m}{\hbox{ m}}

\newcommand{\eqn}[1]{(\ref{#1})}

\newcommand{\lag}{\ensuremath{\mathcal{L}}}
\newcommand{\M}{\ensuremath{\mathcal{M}}}
\newcommand{\D}{\ensuremath{\mathcal{D}}}
\def\tr#1{\mathrm{tr}#1}

\newcommand{\ewgg}{\ensuremath{SU(2)_L \otimes U(1)_Y}}
\def\ws{$SU(2)_L\otimes U(1)_Y$}

\def\ltap{\mathop{\raisebox{-.4ex}{\rlap{$\sim$}} 
\raisebox{.4ex}{$<$}}}
\def\gtap{\mathop{\raisebox{-.4ex}{\rlap{$\sim$}} 
\raisebox{.4ex}{$>$}}}

\def\bentarrow{\:\raisebox{1.3ex}{\rlap{$\vert$}}\!\rightarrow}

\def\dk#1#2#3{
	\begin{equation}
	\begin{array}{r c l}
	#1 & \rightarrow & #2 \\
	 & & \bentarrow #3
	\end{array}
	\end{equation}
		}

\def\dknuc#1#2#3{
	\begin{equation}
	\begin{array}{r c l}
	#1 & \rightarrow & #2 \\
	 & & \phantom{^{152}\;}\bentarrow #3
	\end{array}
	\end{equation}
		}

\newcommand{\cfrac}[2]{\textstyle \frac{#1}{#2}}

\def\vev#1{\langle #1\rangle_0}
\def\abs#1{\left| #1\right|}

\def\bentarrow{\:\raisebox{1.3ex}{\rlap{$\vert$}}\!\rightarrow}

\def\slashi#1{\rlap{\sl/}#1}
\def\slashii#1{\setbox0=\hbox{$#1$}             % set a box for #1
   \dimen0=\wd0                                 % and get its size
   \setbox1=\hbox{\sl/} \dimen1=\wd1            % get size of /
   \ifdim\dimen0>\dimen1                        % #1 is bigger
      \rlap{\hbox to \dimen0{\hfil\sl/\hfil}}   % so center / in box
      #1                                        % and print #1
   \else                                        % / is bigger
      \rlap{\hbox to \dimen1{\hfil$#1$\hfil}}   % so center #1
      \hbox{\sl/}                               % and print /
   \fi}                                         %
\def\slashiii#1{\setbox0=\hbox{$#1$}#1\hskip-\wd0\hbox to\wd0{\hss\sl/\/\hss}}
\def\slashiv#1{#1\llap{\sl/}}
\def\onetev{1-TeV scale}

\def\url#1{\mbox{\href{#1}{\sf #1}}}
\def\urll#1#2{\mbox{\href{#1}{\sf #2}}}

\newcommand{\hepex}[1]{hep-ex/#1}
\newcommand{\hepph}[1]{hep-ph/#1}
\newcommand{\hepth}[1]{hep-th/#1}
\newcommand{\nuclth}[1]{nucl-th/#1}
\newcommand{\heplat}[1]{hep-lat/#1}
\newcommand{\astro}[1]{astro-ph/#1}
\def\prl#1#2#3{\frenchspacing{\it Phys. Rev. Lett. }{\bf #1}, #2 (19#3)}
\def\prll#1#2#3{\frenchspacing{\it Phys. Rev. Lett. }{\bf #1}, #2 (#3)}
\def\pr#1#2#3{\frenchspacing{\it Phys. Rev. D}{\bf #1}, #2 (19#3)}
\def\prev#1#2#3{\frenchspacing{\it Phys. Rev. }{\bf #1}, #2 (19#3)}
\def\prM#1#2#3{\frenchspacing{\it Phys. Rev. D}{\bf #1}, #2 (#3)}

\def\pl#1#2#3{\frenchspacing{\it Phys. Lett. }{\bf #1}, #2 (19#3)}
\def\np#1#2#3{\frenchspacing{\it Nucl. Phys. }{\bf #1}, #2 (19#3)}
\def\npM#1#2#3{\frenchspacing{\it Nucl. Phys. }{\bf #1}, #2 (#3)}
\def\rmp#1#2#3{\frenchspacing{\it Rev. Mod. Phys. }{\bf #1}, #2 (19#3)}
\def\ajp#1#2#3{\frenchspacing{\it Am. J. Phys. }{\bf #1}, #2 (19#3)}

\def\prep#1#2#3{\frenchspacing{\it Phys. Rep. }{\bf #1}, #2 (19#3)}
\def\arnps#1#2#3{\frenchspacing{\it Ann. Rev. Nucl. Part. Sci. }{\bf #1}, #2 (19#3)}
\def\arnpsM#1#2#3{\frenchspacing{\it Ann. Rev. Nucl. Part. Sci. }{\bf #1}, #2 (#3)}
\def\ib#1#2#3{{\bf #1}, #2 (19#3)}
\def\app#1#2#3{\frenchspacing{\it Acta Phys. Polon. B}{\bf #1}, #2 (19#3)}
\def\npbps#1#2#3{{\em Nucl. Phys. B (Proc. Supp.)\/} {\bf #1} (19#3) #2}
\def\phystoday#1#2#3#4{\frenchspacing{\it Phys. Today\/ }{\bf #1}, #2 
(\ifcase#3\or January\or 
         February\or March\or April\or May\or June\or July\or August\or 
         September\or October\or November\or December\fi, 19#4)}
\def\uspek#1#2#3#4#5#6{{\it Usp. Fiz. Nauk }{\bf #1}, #2 (19#3) [English translation: 
         {\it Sov. Phys.--Uspekhi }{\bf #4}, #5 (19#6)]}

\begin{document}

\title{The Electroweak Theory}

\author{Chris Quigg}

\address{Theoretical Physics Department \\ Fermi National Accelerator
Laboratory \\ Batavia, IL 60510 USA\\
\urll{mailto:quigg@fnal.gov}{E-mail: quigg@fnal.gov}}


\maketitle

\abstracts{
After a short essay on the current state of particle physics, I review
the antecedents of the modern picture of the weak and electromagnetic
interactions and then undertake a brief survey of the
$SU(2)_{L}\otimes U(1)_{Y}$ electroweak theory.  I review the features
of electroweak phenomenology at tree level and beyond, present an
introduction to the Higgs boson and the \onetev, and examine arguments
for enlarging the electroweak theory.  I conclude with a brief look at
low-scale gravity.
}
\section{Introduction}

\subsection{Our picture of matter}
At the turn of the third millennium, we base our 
understanding of physical phenomena on the identification of a few 
constituents that seem elementary at the current limits of resolution 
of about $10^{-18}\m$, and a few fundamental forces.  The 
constituents are the pointlike quarks
	\begin{equation}
		\left(
		\begin{array}{c}
			u  \\
			d
		\end{array}
		 \right)_{L} \;\;\;\;\;\;
		\left(
		\begin{array}{c}
			c  \\
			s
		\end{array}
		 \right)_{L} \;\;\;\;\;\;
		\left(
		\begin{array}{c}
			t  \\
			b
		\end{array}
		 \right)_{L}\; ,		 
	\end{equation}
and leptons
		\begin{equation}
		\left(
		\begin{array}{c}
			\nu_{e}  \\
			e^{-}
		\end{array}
		 \right)_{L} \;\;\;\;\;\;
		\left(
		\begin{array}{c}
			\nu_{\mu}  \\
			\mu^{-}
		\end{array}
		 \right)_{L} \;\;\;\;\;\;
		\left(
		\begin{array}{c}
			\nu_{\tau}  \\
			\tau^{-}
		\end{array}
		 \right)_{L}\; ,		 
	\end{equation}
with strong, weak, and electromagnetic interactions specified by
$SU(3)_{c}\otimes SU(2)_{L}\otimes U(1)_{Y}$ gauge symmetries.

This concise statement of the standard model invites us to consider 
the agenda of particle physics today under four themes.  
\textit{Elementarity.} Are the quarks and leptons structureless, or 
will we find that they are composite particles with internal 
structures that help us understand the properties of the individual 
quarks and leptons?  \textit{Symmetry.} One of the most powerful 
lessons of the modern synthesis of particle physics is that 
symmetries prescribe interactions.  Our investigation of symmetry must 
address the question of which gauge symmetries exist (and, eventually, 
why).  We must also understand how the electroweak 
symmetry\footnote{and, no doubt, others---including the symmetry that 
brings together the strong, weak, and electromagnetic interactions.} 
is hidden.  The most urgent problem in particle physics is to complete 
our understanding of electroweak symmetry breaking by exploring the 
1-TeV scale.  \textit{Unity.} We have the fascinating possibility of 
gauge coupling unification, the idea that all the interactions we 
encounter have a common origin---and thus a common strength---at suitably 
high energy.  Next comes the imperative of anomaly freedom in the 
electroweak theory, which urges us to treat quarks and leptons 
together, not as completely independent species.  Both ideas are 
embodied in unified theories of the strong, weak, and electromagnetic 
interactions, which imply the existence of still other forces---to 
complete the grander gauge group of the unified theory---including 
interactions that change quarks into leptons.  Supersymmetry and the self-interacting 
quanta of non-Abelian theories both hint that the 
traditional distinction between force particles and constituents might 
give way to a unified understanding of all the particles.  
\textit{Identity.} We do not understand the physics that sets quark 
masses and mixings.  Although experiments are testing the idea that the phase 
in the quark-mixing matrix lies behind the observed \textsf{CP} 
violation, we do not know what determines that phase.  The 
accumulating evidence for neutrino oscillations presents us with a new 
embodiment of these puzzles in the lepton sector.  At bottom, the 
question of identity is very simple to state: What makes an electron 
an electron, a neutrino a neutrino, and a top quark a top quark?

\subsection{QCD is part of the standard model}
The quark model of hadron structure and the parton model of 
hard-scattering processes have such pervasive influence on the way we 
conceptualize particle physics that quantum chromodynamics, the theory 
of strong interactions that underlies both, often fades into the 
background when the standard model is discussed.  I want to begin 
these 
lectures on the electroweak theory with a clear statement that QCD 
is indeed part of the standard model, and with the belief that 
understanding QCD may be indispensable for deepening our understanding 
of the electroweak theory.  Other lecturers will explore the 
application of QCD to flavor physics.

Quantum chromodynamics is a remarkably simple, successful, and rich 
theory of the strong interactions.\footnote{For a passionate elaboration 
of this statement, see Frank Wilczek's keynote address at PANIC '99, 
Ref.\ \cite{fw}.  An authoritative portrait of QCD and its many applications 
appears in the monograph by Ellis, Stirling, and Webber, Ref.\ \cite{esw}.}  The 
perturbative regime of QCD exists, thanks to the crucial property of 
asymptotic freedom, and describes many phenomena in quantitative 
detail.  The strong-coupling regime controls hadron structure and 
gives us our best information about quark masses.

\begin{figure}[tb] 
\centerline{\BoxedEPSF{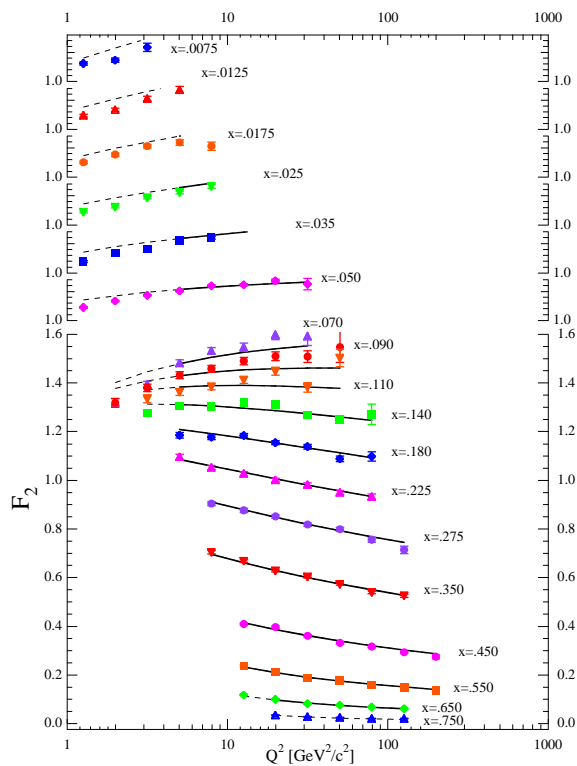 scaled 530}}
\vspace{10pt}
\caption{The structure function $F_2$ measured in $\nu N$ interactions,
from Ref.\ {\protect\cite{csb}}.}
\label{fig:F2}
\end{figure}

The classic test of perturbative QCD is the prediction of subtle 
violations of Bjorken scaling in deeply inelastic lepton scattering.  
As an illustration of the current state of the comparison between 
theory and experiment, I show in Figure \ref{fig:F2} the singlet 
structure function $F_{2}(x,Q^{2})$ measured in $\nu N$ 
charged-current interactions by the CCFR Collaboration at Fermilab.  
The solid lines for $Q^{2} \gtap (5\gevc)^{2}$ represent QCD fits; 
the dashed lines extrapolate to smaller values of $Q^{2}$.  As 
we see in this example, modern data are so precise that one can search 
for small departures from the QCD expectation.

\begin{figure}[tb] 
\centerline{\BoxedEPSF{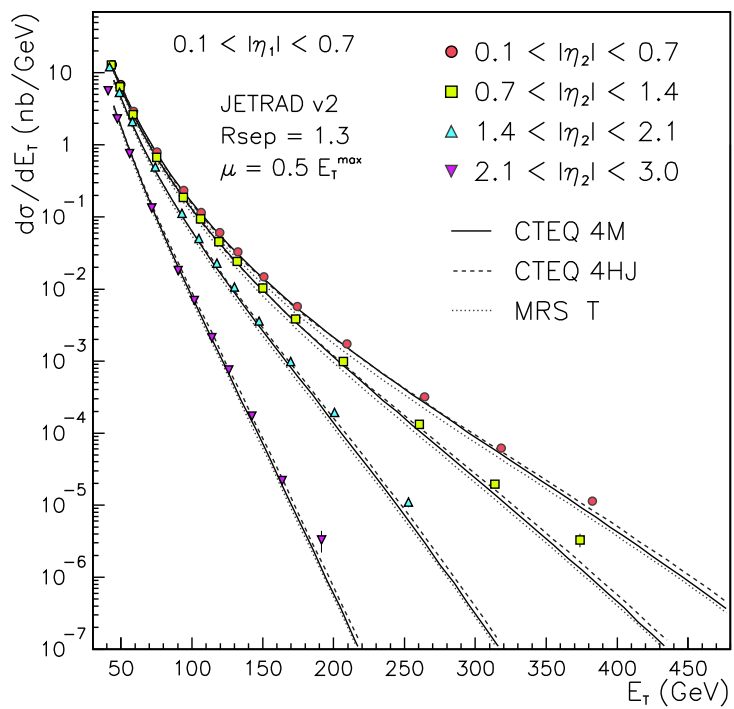 scaled 530}}
\vspace{10pt}
\caption{Cross sections measured at $\sqrt{s} = 1.8\tev$ by the CDF 
Collaboration for central jets (defined by $0.1 < \abs{\eta_{1}} < 
0.7$), with the second jet confined to specified intervals in the 
pseudorapidity $\eta_{2}$.{\protect\cite{cdfjets}}  The curves show 
next-to-leading-order QCD predictions based on the CTEQ4M (solid 
line), CTEQ4HJ (dashed line), and MRST (dotted line) parton 
distributions.}
\label{fig:jets}
\end{figure}

Perturbative QCD also makes spectacularly successful predictions for 
hadronic processes.  I show in Figure \ref{fig:jets} that pQCD, 
evaluated at next-to-leading order using the program \textsc{jetrad,} 
accounts for the transverse-energy spectrum of central jets produced 
in the reaction
\begin{equation}
    \bar{p}p \to \hbox{jet}_{1} + \hbox{jet}_{2} + \hbox{anything}
    \label{eq:pbarpjj}
\end{equation}
over at least six orders of magnitude, at $\sqrt{s} = 
1.8\tev$.\footnote{For a systematic review of high-$E_{T}$ jet 
production, see Blazey and Flaugher, Ref.\ \cite{brenna}.}

%%%%%%%%%%%%%%%%%%%%%%%%%%%%%%%%%%%%%%%%%%%%%%%%%%%%%%%%%%%%%%%%%%%%%%
%                                                                    %
%   The running of the strong coupling constant in CDF jets \ldots   %
%   Figure \ref{fig:CDFalpha} \ldots                                 %
%   \begin{figure}[tb]                                               %
%   \centerline{\BoxedEPSF{CDFRunningalphax.eps scaled 600}}          %
%   \vspace{10pt}                                                    %
%   \caption{Running $\alpha_{s}$ from CDF \ldots}                   %
%   \label{fig:CDFalpha}                                             %
%   \end{figure}                                                     %
%                                                                    %
%%%%%%%%%%%%%%%%%%%%%%%%%%%%%%%%%%%%%%%%%%%%%%%%%%%%%%%%%%%%%%%%%%%%%%

\begin{figure}[tb] 
\centerline{\BoxedEPSF{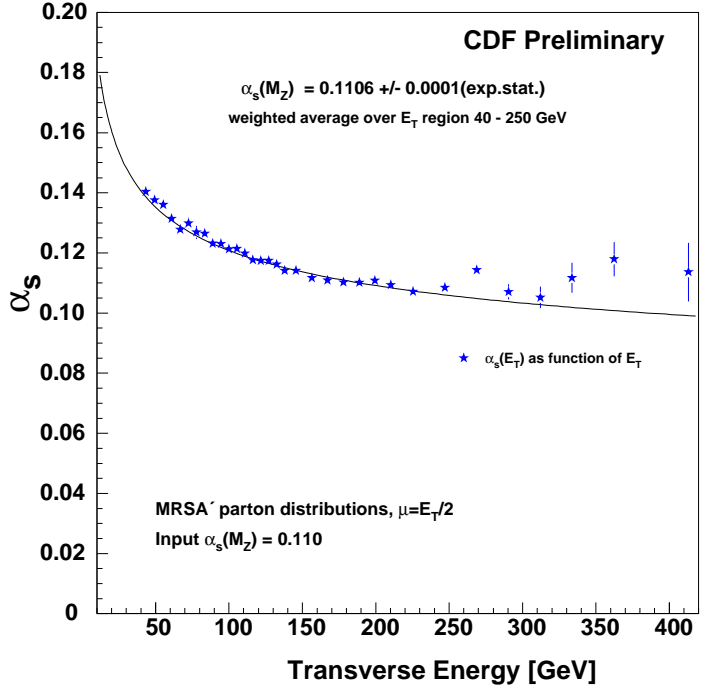 scaled 500}}
\vspace{10pt}
\caption{Determinations of $\alpha_{s}$ inferred from the comparison
of measured inclusive jet cross sections with the \textsc{jetrad} NLO
Monte-Carlo program.  Source of this figure
is {\protect\urll{http://www-cdf.fnal.gov/physics/new/qcd/qcd99_pub_blessed.html}
{http://www-cdf.fnal.gov/physics/new/qcd/qcd99\_pub\_blessed.html}}.}
\label{fig:CDFalpha}
\end{figure}

The $Q^{2}$-evolution of the strong coupling constant predicted by 
QCD, which in lowest order is
\begin{equation}
    1/\alpha_{s}(Q^{2}) = 1/\alpha_{s}(\mu^{2}) +
    \frac{33-2n_{f}}{12\pi} \log(Q^{2}/\mu^{2}) \; ,
    \label{eq:runalph}
\end{equation}
where $n_{f}$ is the number of active quark flavors, has been observed
within individual experiments\cite{cdfrun,leprun} and by comparing
determinations made in different experiments at different 
scales.\footnote{For a review, see Hinchliffe and Manohar, Ref.\
\cite{ianeesh}.} 
Figure \ref{fig:CDFalpha}, from the CDF Collaboration, shows the 
values of $\alpha_{s}(E_{T})$ inferred from jet production cross sections in 
1.8-TeV $\bar{p}p$ collisions.  The curve shows the expected running 
of the strong coupling constant. 

\begin{figure}[tb] 
\centerline{\BoxedEPSF{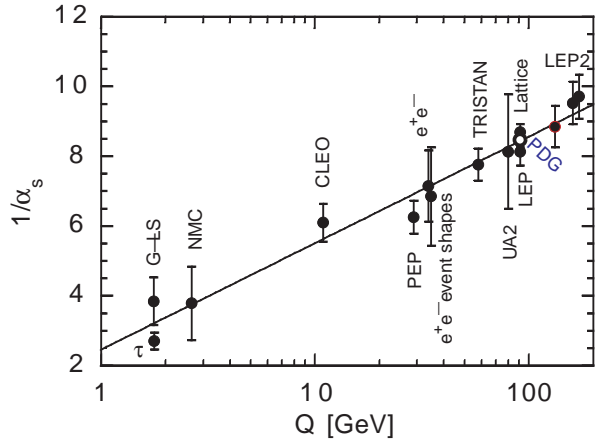 scaled 700}}
\vspace{10pt}
\caption{Determinations of $1/\alpha_{s}$, plotted at the scale 
$\mu$ at which the measurements were made.  The line shows the 
expected evolution {\protect\eqn{eq:runalph}}.}
\label{fig:oneoveralpha}
\end{figure}
\begin{figure}[b!] 
\centerline{\BoxedEPSF{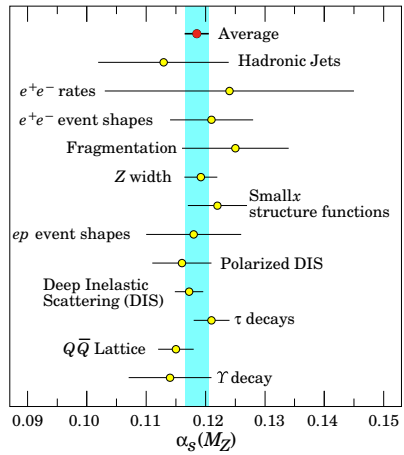 scaled 300}}
\vspace{10pt}
\caption{Determinations of $\alpha_{s}(M_{Z})$ from several 
processes. In most cases, the value measured at a scale $\mu$ has been 
evolved to $\mu = M_{Z}$.  Error bars include the theoretical uncertainties.
From the \textit{Review of Particle Physics,} Ref.\ {\protect\cite{pdg}}.}
\label{fig:asMZ}
\end{figure}
A compilation of
$1/\alpha_{s}$ determinations from many experiments, shown in Figure
\ref{fig:oneoveralpha}, exhibits the expected behavior.\footnote{A
useful plot of $\alpha_{s}$ \textit{vs.} $Q^{2}$ appears as Fig.~9.2
of the \textit{Review of Particle Physics,} Ref.\ \cite{pdg}.}
Evolved to a common scale $\mu = M_{Z}$, the various 
determinations of $\alpha_{s}$ lead to consistent values shown in 
Figure \ref{fig:asMZ}.
%\clearpage
\subsection{Sources of mass}
We sometimes hear the statement that the discovery of the Higgs boson 
will reveal the origin of all mass.  I am the first to say that 
unraveling the origin of electroweak symmetry breaking---for which 
``the discovery of the Higgs boson'' is a common shorthand---will be a 
spectacular achievement in the history of science, but we are 
physicists, and we should say what we mean.  There are, in fact, 
several sources of mass, and we can imagine soon understanding them 
all.  At a level we now find so commonplace as to seem trivial, we 
understand the mass of any atom or molecule in terms of the masses of 
the atomic nuclei, the mass of the electron, and quantum 
electrodynamics.\footnote{This is the sense in which quantum theory 
explains all of chemistry.  The calculations are hard enough that we 
leave them to chemists!}  And in precise and practical---if not quite 
``first-principle''---terms, we understand the masses of all the 
nuclides in terms of the proton mass, the neutron mass, and our 
knowledge of nuclear forces.  

\begin{figure}[tb] 
\centerline{\BoxedEPSF{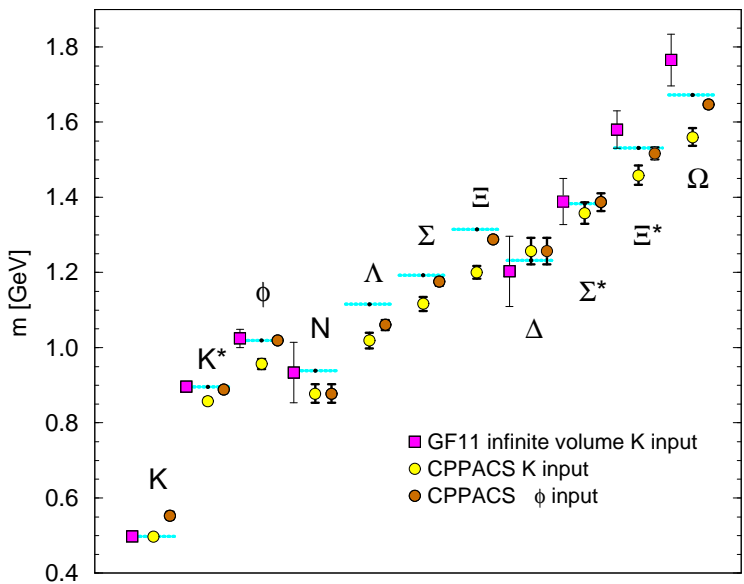 scaled 600}}
\vspace{10pt}
\caption{Final results of the CP-PACS Collaboration's quenched light 
hadron spectrum in the continuum limit.{\protect\cite{burk}}  
Experimental values (horizontal lines) and earlier results from the 
GF11 Collaboration{\protect\cite{ref:GF11}} are plotted for 
comparison.}
\label{fig:hadrons}
\end{figure}

What about the proton and neutron 
masses?\footnote{An accessible essay on our understanding of hadron 
mass appears in Ref.\ \cite{fwpt}.}  Do we require the Higgs mechanism to understand 
them?\footnote{The standard model of particle physics has taught us 
many fascinating interrelations, including the effect of heavy-quark 
masses on the low-energy value of $\alpha_{s}$, which sets the scale 
of the light-hadron masses.  For a quick tour, see my \textit{Physics Today} 
article on the top quark\cite{cqpt}; Bob Cahn's \textit{RMP} 
Colloquium\cite{rncrmp} is a more expansive tour of connections in the standard 
model.}
Thanks to QCD, we have learned that the dominant contribution to the 
light-hadron masses is not the masses of the quarks of which they are 
constituted, but the energy stored up in confining the quarks in a 
tiny volume. Our most useful tool in the 
strong-coupling regime is lattice QCD.  Calculating the light-hadron 
spectrum from first principles has been one of the main objectives of 
the lattice program, and important strides have been made recently.  
In 1994, the GF11 Collaboration\cite{ref:GF11} carried out a quenched 
calculation of the spectrum (no dynamical fermions) that yielded 
masses that agree with experiment within 5--10\%, with good 
understanding of the residual systematic uncertainties.  The CP-PACS 
Collaboration centered in Tsukuba has embarked on an ambitious program 
that will soon lead to a full (unquenched) calculation.  Their 
quenched results, along with those of the GF11 Collaboration, are 
presented in Figure \ref{fig:hadrons}.\cite{burk}  The gross features 
of the light-hadron spectrum are reproduced very well, but if you look 
with a critical eye (as the CP-PACS collaborators do), you will notice
that the quenched light hadron spectrum systematically deviates from 
experiment.  The $K$-$K^*$ mass splitting is underestimated by about 
$10\%$, and the results differ depending on whether the 
strange-quark mass is fixed from the $K$ mass or the $\phi$ mass.  The 
forthcoming unquenched results should improve the situation further, 
and give us new insights into how well---and why!---the simple quark 
model works.

We also have a reasonably good understanding of the masses of the 
electroweak gauge bosons, as we will develop in \S\ref{sec:IVB}.\cite{fwpt2}
Gauge-boson masses are predicted in terms of the gauge coupling $g$ 
and the weak 
mixing parameter $\sin^{2}\theta_{W}$:
\begin{eqnarray}
	M_{W}^{2} & = & \frac{g^{2}v^{2}}{2} = 
	\frac{\pi\alpha}{G_{F}\sqrt{2}\sin^{2}\theta_{W}} \; ,
	\label{gbmsw}  \\
	M_{Z}^{2} & = & \frac{M_{W}^{2}}{\cos^{2}\theta_{W}} \; ,
	\label{gbmsz}
\end{eqnarray}
where $v = (G_{F}\sqrt{2})^{-1/2} = 246\gev$ sets the electroweak 
scale.  Completing our understanding of the mechanism that endows the 
gauge bosons with mass is what we expect to accomplish by exploring 
the \onetev; that is what we can promise that the discovery of the 
Higgs boson---broadly understood---will deliver.
While we don't yet have a complete understanding of the 
electroweak scale or the value of the weak mixing parameter, we can 
imagine how those two quantities might arise in unified theories or 
from new (strong) dynamics.

The masses of the elementary fermions are a more mysterious 
story:
Each fermion mass involves a new, so far incalculable, 
Yukawa coupling.  For example, the term in the electroweak Lagrangian 
that gives rise to the electron mass is
\begin{equation}
	\lag_{\mathrm{Yuk}} = -\zeta_{e}\left[\bar{\mathsf{R}}(\varphi^{\dagger}\mathsf{L}) + 
	(\bar{\mathsf{L}}\varphi)\mathsf{R}\right]\; ,
	\label{eyuk}
\end{equation}
where $\varphi$ is the (complex) Higgs field and the left-handed and 
right-handed fermions are specified as
\begin{equation}
	\mathsf{L} = \left(
	\begin{array}{c}
		\nu_{e}  \\
		e
	\end{array}
	\right)_{\mathrm{L}}\; , \quad \mathsf{R} = e_{\mathrm{R}}
	\label{leptons}
\end{equation}
When the electroweak symmetry is spontaneously broken, the electron 
mass emerges as
\begin{equation}
	m_{e} = \zeta_{e}v/\sqrt{2}\; .
	\label{emass}
\end{equation}
The Yukawa couplings that reproduce the observed quark and lepton 
masses range over many orders of magnitude, from $\zeta_{e} \approx 
3 \times 10^{-6}$ for the electron to $\zeta_{t} \approx 1$ for the 
top quark.  Their origin is unknown.

In one sense, therefore, \textit{all fermion masses involve physics
beyond the standard model.}\cite{ggrtasi} We cannot be sure that finding the Higgs
boson, or understanding electroweak symmetry breaking, will bring
enlightenment about the fermion masses.  Neutrino masses may have a
different origin than the masses of the quarks and charged leptons:
alone among the known fermions, the neutral neutrino can be its own
antiparticle.\cite{boris2K}  This fact opens the possibility of several varieties of
neutrino masses.

It is worth remarking on another manifestation of the logical 
separation between the origin of gauge-boson masses and the origin of 
fermion masses.  The observation that a fermion mass is different from 
zero ($m_{f} \neq 0$) implies that the electroweak gauge symmetry 
\ewgg\ is broken, but electroweak symmetry breaking is only a 
necessary, not a sufficient, condition for the generation of fermion 
mass.  The separation is complete in simple technicolor,\cite{TC} the 
theory of dynamical symmetry breaking modeled on the 
Bardeen--Cooper--Schrieffer theory of the superconducting phase 
transition.

Finally, the electroweak theory we are about to describe does not 
predict the mass of the Higgs boson, and there is no assurance that 
finding the Higgs boson will tell us the origin of this mass.

Will the discovery of the Higgs boson be stupendously important? 
Beyond any doubt, as the rest of these lectures will begin to show. 
But will it explain \textit{the origin of all mass?} Be careful what
you promise!

\section{Antecedents of the Electroweak Theory}
In \textit{The Odd Quantum,} Sam 
Treiman\cite{sbt} quotes from the 1898--99 University of Chicago 
catalogue: ``While it is never safe to affirm that the future of the 
Physical Sciences has no marvels in store even more astonishing than 
those of the past, it seems probable that most of the grand underlying 
principles have been firmly established and that further advances are 
to be sought chiefly in the rigorous application of these principles 
to all the phenomena which come under our notice \ldots .  An eminent 
physicist has remarked that the future truths of Physical Science are 
to be looked for in the sixth place of decimals.''  These confident 
words were written, we now know, just as the classical world of 
determinism and uncuttable atoms and continuous distributions of 
energy was beginning to come apart.  Much crucial progress did come 
from precise measurements---not always in the sixth place of the 
decimals, but precise nonetheless.  At the same time, the century we are 
leaving has repeatedly shown that Nature's marvels are not 
limited by our imagination, and that exploration can yield surprises 
that completely change the way we think.

Before we leap into a discussion of the modern electroweak theory, it 
will be useful to spend a few moments recalling the soil in which the 
electroweak theory grew.  We shall not attempt anything resembling a 
full intellectual history,\cite{bram,Fitch:1994cq,combuck,cahngold}  but only hit a few of the high 
spots.
\subsection{Radioactivity, $\beta$ decay, and the neutrino}
Becquerel's discovery of radioactivity in 1896 is one of the
wellsprings of modern physics.  In a short time, physicists learned 
to distinguish several sorts of radioactivity, classified by 
Rutherford according to the energetic projectile emitted in the spontaneous 
disintegration.  Natural and artificial radioactivity includes nuclear 
$\beta$ decay, observed as
\begin{equation}
	^{A}{\mathrm{Z}} \to\ ^{A}({\mathrm{Z+1}}) + \beta^{-}\; ,
	\label{eq:betadk}
\end{equation}
where $\beta^{-}$ is Rutherford's name for what was soon identified 
as the electron and
$^{A}{\mathrm{Z}}$ stands for the nucleus with $Z$ protons and $A-Z$
neutrons.  Examples\footnote{It is a curious fact that 
$\beta^{+}$-emitters, $^{A}{\mathrm{Z}} \to\ ^{A}({\mathrm{Z-1}}) + 
\beta^{+}$, are rare among the naturally occurring isotopes.  The 
first example, radio-phosphorus produced in $\alpha \mathrm{Al}$ 
collisions, was found by Ir\`{e}ne and Fr\'{e}d\'{e}ric Joliot-Curie in
1934, after the 
discovery of the positron in cosmic rays.  In our time, the decay 
$^{19}{\mathrm{Ne}} \to\ ^{19}{\mathrm{F}} + \beta^{+}$ has been a 
favorite for the study of right-handed charged currents and time 
reversal invariance.  And, of course, the positron emitters form the 
technological basis for positron-emission tomography.}
are tritium $\beta$ decay,
\begin{equation}
	^{3}\mathrm{H}_{1} \to\ ^{3}\mathrm{He}_{2} + \beta^{-}\; ,
\end{equation}
neutron $\beta$ decay,
\begin{equation}
	n \to p + \beta^{-}\; ,
\end{equation}
and $\beta$ decay of Lead-214,
\begin{equation}
	^{214}\mathrm{Pb}_{82} \to\ ^{214}\mathrm{Bi}_{83} + \beta^{-}\; .
\end{equation}

\begin{figure}[tb] 
\centerline{\BoxedEPSF{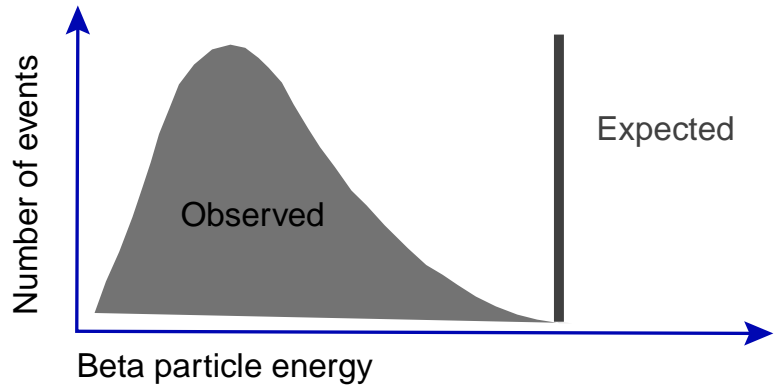  scaled 1250}}
\vspace{10pt}
\caption{Expectations and reality for the beta decay spectrum.}
\label{fig:betaspec}
\end{figure}

For  two-body decays, the Principle of Conservation of Energy \&
Momentum says that the $\beta$ particle should have a definite energy, 
indicated by the spike in Figure \ref{fig:betaspec}.  What was 
observed was very different: in 1914, James Chadwick (later to
discover the neutron) showed conclusively\cite{chadwick} that in the
decay of Radium B and C ($^{214}$Pb and $^{214}$Bi), the 
$\beta$ energy follows a continuous spectrum, as shown in Figure
\ref{fig:betaspec}.

What could be the meaning of this completely unexpected behavior?  
Niels Bohr was willing to consider the possibility
that energy and momentum are not uniformly conserved in 
subatomic events.  The $\beta$-decay energy crisis tormented physicists for years.  On 
December 4, 1930, Wolfgang Pauli addressed an open letter\footnote{Pauli's letter (in the 
original German) is reproduced in Ref.\ \cite{pauliworks}. For an 
English translation, see pp.~127-8 of Ref.\ \cite{bram}.  It begins, 
``Dear Radioactive Ladies and Gentlemen, I have hit upon a desperate 
remedy regarding \ldots the continuous $\beta$-spectrum \ldots'' Pauli 
concluded, ``For the moment I dare not publish anything about this 
idea and address myself confidentially first to you.  \ldots I admit 
that my way out may seem rather improbable \textit{a priori} \ldots .  
Nevertheless, if you don't play you can't win \ldots .  Therefore, 
Dear Radioactives, test and judge.'' Pauli's neutrino, together with 
the discovery of the neutron, also resolved a vexing nuclear 
spin-and-statistic problem.} to a meeting on radioactivity in 
T\"{u}bingen.  Pauli could not attend in person because his presence 
at a student ball in Z\"{u}rich was ``indispensable.'' In his letter, 
Pauli advanced the outlandish idea of a new, very penetrating, neutral 
particle of vanishingly small mass.  Because Pauli's new particle 
interacted very feebly with matter, it would escape undetected from 
any known apparatus, taking with it some energy, which would seemingly 
be lost.  The balance of energy and momentum would be restored by the
particle we now know as the electron's antineutrino.  The proper 
scheme for beta decay is thus
\begin{equation}
	^{A}{\mathrm{Z}} \to\ ^{A}({\mathrm{Z+1}}) + \beta^{-} + \bar{\nu} \; .
	\label{eq:betadknu}
\end{equation}

Pauli's new particle was indeed a ``desperate remedy,'' but it was, 
in its way, very conservative, for it preserved the principle of 
energy and momentum conservation and with it the notion that the laws 
of physics are invariant under translations in space and time.  The
hypothesis fit the facts.\footnote{As you TASI students continue in
physics, you will be amazed and delighted to find how quickly we
learn---or how little we knew just a short time ago.  It is easy to
assume that anything we read in textbooks has been known forever, so
it is sometimes stunning to learn that something we take for granted
actually had to be discovered.  An example that has both scientific
and touristic interest for students in Boulder is Jack Steinberger's
discovery on Mount Evans of the continuous electron spectrum in muon
decay.{\protect\cite{mrjack}}} After Chadwick's discovery of the neutron in
1932, Fermi named Pauli's hypothetical particle the neutrino, to distinguish it
from the neutron, and constructed his four-fermion theory of the weak
interaction.  Experimental confirmation of Pauli's neutrino had to
wait for dramatic advances in technology.\footnote{Detecting a
particle as penetrating as the neutrino required a large target and a
copious source of neutrinos.  In 1953, Clyde Cowan and Fred Reines\cite{cowan} 
used the intense beam of antineutrinos from a fission
reactor $^{A}{\mathrm{Z}} \to\ ^{A}({\mathrm{Z+1}}) + \beta^{-} +
\bar{\nu}\;$, and a heavy target ($10.7~\mathrm{ft}^{3}$ of liquid
scintillator) containing about $10^{28}$ protons to detect the
reaction $\bar{\nu} + p \to\ e^{+} + n$.  Initial runs at the Hanford
Engineering Works were suggestive but inconclusive.  Moving their
apparatus to the stronger fission neutrino source at the Savannah
River nuclear plant, Cowan and Reines and their team made the
definitive observation of inverse $\beta$ decay in 1956.\cite{reines}}
 
We can now recognize $\beta$ decay as the first hint for \textit{flavor,} the 
theme of this summer school.  Indeed, neutron beta decay is the 
prototype charged-current, flavor-changing interaction:
    \begin{center} \begin{picture}(335,50)(0,0)
	\ArrowArc(150,25)(15,20,90)
	\ArrowArcn(150,25)(15,-20,-90)
	\Text(171,25)[c]{$W$}
	\Text(145,40)[]{$p$}
	\Text(145,10)[]{$n$}
	\ArrowArcn(190,25)(15,160,90)
	\ArrowArc(190,25)(15,-160,-90)
	\Text(195,40)[]{$e$}
	\Text(195,10)[]{$\bar{\nu}$}
	\Text(205,25)[c]{$\cdot$}
    \end{picture}   \end{center}
%\clearpage
\subsection{The neutron and flavor symmetry}
The discovery of the neutron made manifest the case for flavor, with 
two species of nucleon nearly degenerate in mass:
\begin{eqnarray}
    M(n) & = & 939.565\,63 \pm 0.000\,28\mevcc
    \nonumber  \\
    M(p) & = & 938.272\,31 \pm 0.000\,28\mevcc
    \label{eq:nucleons}  \\
    \Delta M & =  & 1.293318 \pm 0.000\,009\mevcc
    \nonumber
\end{eqnarray}
so that $\Delta M/M \approx 1.4 \times 10^{-3}$.  The similarity of 
the neutron and proton masses makes it plausible to inquire into the 
charge independence of nuclear forces.  The two-nucleon system is not 
particularly informative: among $NN$ states, $pp$ and $nn$ are 
unbound, and only the isocalar $np$ state, the deuteron, is very 
lightly bound.  Hints for the charge independence of nuclear forces 
come from many light nuclei.  We may compare the binding energy of 
\begin{eqnarray}
    ^{3}\mathrm{H}(ppn) & = & 8.481\,855 \pm 0.000\,013\mev
    \nonumber  \\
    ^{3}\mathrm{He}(pnn) & = & 7.718\,109 \pm 0.000\,013\mev
    \label{eq:tritons}  \\
    \Delta(\mathrm{B.E.}) & = & 0.763\,46\mev\; .
    \nonumber
\end{eqnarray}
The difference in binding energy is very close to a primitive 
estimate for the Coulomb repulsion in $^{3}\mathrm{He}$: taking the 
measured charge radius of $r = 1.97 \pm 0.015\fm$, we estimate the 
Coulomb energy as $\alpha/r \approx 0.731\mev$.

\begin{figure}[t!] 
\centerline{\BoxedEPSF{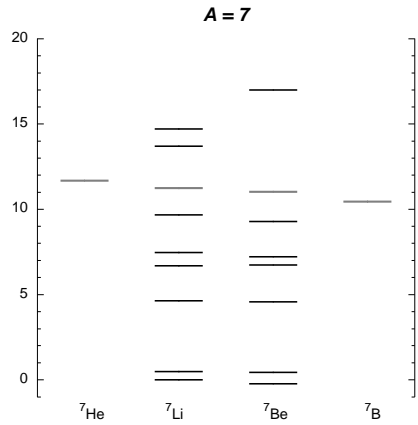 scaled 395}}
\vspace{10pt}
\caption{Simplified isobar diagram for the $A=7$ nuclei.  The 
presumed $I=3/2$ isobaric analogue levels are shown in grey.  
Following the usual practice, the diagrams for individual isobars are 
shifted vertically to eliminate the $n$-$p$ mass difference and the 
Coulomb energy.  Data taken from Ref.~{\protect\cite{ajz7}}.}
\label{fig:A7}
\end{figure}

\begin{figure}[b!] 
\centerline{\BoxedEPSF{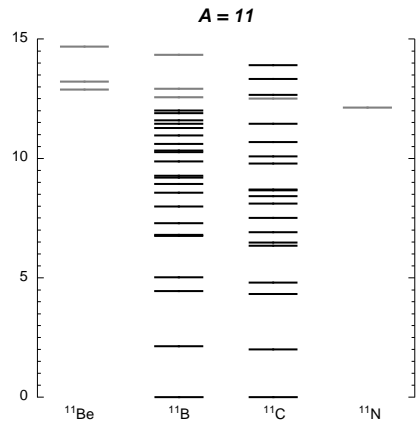 scaled 395}}
\vspace{10pt}
\caption{Simplified isobar diagram for the $A=11$ nuclei.  The 
presumed $I=3/2$ isobaric analogue levels are shown in grey.  
Following the usual practice, the diagrams for individual isobars are 
shifted vertically to eliminate the $n$-$p$ mass difference and the 
Coulomb energy.  Data taken from Ref.~{\protect\cite{ajz11}}.}
\label{fig:A11}
\end{figure}

More detailed evidence that nuclear forces are the same for protons 
and neutrons come from the level structures in mirror nuclei.  I show 
in Figures \ref{fig:A7} and \ref{fig:A11} the kinship between the 
$^{7}\mathrm{Li}(4p+3n)$ and $^{7}\mathrm{Be}(3p+4n)$ level schemes, 
and between the $^{11}\mathrm{B}(5p+6n)$ and $^{11}\mathrm{C}(6p+5n)$ 
level schemes.  In both cases, isospin $I=3/2$ isobaric analogue 
levels are present in the $^{7}\mathrm{He}(2p+5n)$ and 
$^{7}\mathrm{B}(5p+2n)$ ground states, and the 
$^{11}\mathrm{Be}(4p+7n)$ and $^{11}\mathrm{N}(7p+4n)$ ground states.

\begin{figure}[tb] 
\centerline{\BoxedEPSF{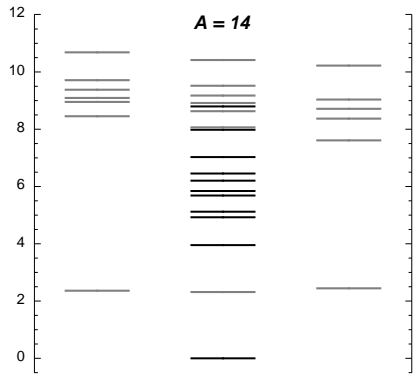 scaled 395}}
\vspace{10pt}
\caption{Simplified isobar diagram for the $A=14$ nuclei.  The 
presumed $I=1$ isobaric analogue levels are shown in grey.  
Following the usual practice, the diagrams for individual isobars are 
shifted vertically to eliminate the $n$-$p$ mass difference and the 
Coulomb energy.  Data taken from Ref.~{\protect\cite{ajz14}}.}
\label{fig:A14}
\end{figure}

Extremely compelling evidence for the charge independence of nuclear 
forces comes from the systematic study of two-nucleon states in the 
$A=14$ nuclei, which consist of two nucleons outside a closed core:
\begin{eqnarray}
    ^{14}\mathrm{O}: & ^{12}\mathrm{C} + (pp) & I_{3}=+1
    \nonumber  \\
    ^{14}\mathrm{N}: & ^{12}\mathrm{C} + (pn) & I_{3}=0
    \label{eq:A14}  \\
    ^{14}\mathrm{C}: & ^{12}\mathrm{C} + (nn) & I_{3}=-1
    \nonumber
\end{eqnarray}
We see in Figure \ref{fig:A14} that the $I=1$ levels are common to all 
three elements.  Nitrogen-14 has many additional $I=0$ levels.

The charge independence of nuclear forces led to the first flavor 
symmetry, isospin invariance.  We regard the proton and neutron as two 
states of the nucleon doublet,
\begin{displaymath}
    \left( 
    \begin{array}{c}
        p  \\
        n
    \end{array}
    \right) \; ,
\end{displaymath}
and consider the nuclear interaction to be invariant under rotations
in isospin space.  In the absence of electromagnetism, which supplies 
the distinguishing information that the proton carries an electric 
charge but the neutron does not, it is \textit{a matter of convention} which 
state---or which superposition of states---we name the proton or 
neutron.\footnote{If we lived in a world without electromagnetism, 
how could you determine that there are, in fact, two species of 
nucleons?}  Once we have the notion of isospin invariance in mind, it 
is a small step to the idea of weak isospin, which we infer from the 
fact that the charged-current weak interaction transforms one member 
of the nucleon doublet into another.  The notion of weak-isospin 
families is a cornerstone of our modern electroweak theory. 
%\clearpage
\subsection{Parity violation in weak decays}
A series of observations and analyses through the 1950s led to the
suggestion that the weak interactions did not respect reflection
symmetry, or parity.  In 1956, C. S. Wu and collaborators detected a
correlation between the spin vector $\vec{J}$ of a polarized
$^{60}\mathrm{Co}$ nucleus and the direction $\hat{p}_{e}$ of the
outgoing $\beta$ particle.\cite{Wu:1957my} Since parity inversion
leaves spin, an axial vector, unchanged:
\begin{equation}
    \mathcal{P}: \vec{J} \to \vec{J}\; ,
    \label{eq:PonJ}
\end{equation}
while reversing the electron direction:
\begin{equation}
    \mathcal{P}: \hat{p}_{e} \to - \hat{p}_{e} \; ,
    \label{eq:Ponp}
\end{equation}
the correlation $\vec{J} \cdot \hat{p}_{e}$ is \textit{parity 
violating.}  Detailed analysis of the $^{60}\mathrm{Co}$ result and 
others that came out in quick succession established that the 
charged-current weak interactions are left-handed.  Since parity 
links a left-handed neutrino with a right-handed neutrino, 
    \begin{center} \begin{picture}(300,30)(0,0)
	\Text(150,15)[]{$\mathcal{P}$}
	\Text(160,15)[l]{$\longleftarrow$  $\nu_{R}$}
	\Text(182,15)[]{{\LARGE $/$}}
	\Text(182,15)[]{{\LARGE $\backslash$}}
	\Text(169,22)[]{$\Leftarrow$}
	\Text(140,15)[r]{$\nu_{L} \longrightarrow$}
	\Text(131,22)[]{$\Leftarrow$}
    \end{picture}   \end{center}
we build a manifestly parity-violating theory with only $\nu_{L}$.
    
How can we establish that the known neutrino is left-handed?  The 
simplest experiment to describe---though not the earliest to measure 
neutrino helicity---is to measure the helicity of the outgoing 
$\mu^{+}$ in the decay of a spin-zero $\pi^{+} \to 
\mu^{+}\nu_{\mu}$.\cite{bardon,possoz}  
    \begin{center} \begin{picture}(300,30)(0,0)
	\BCirc(150,15){10}
	\Text(150,17)[]{$\pi^{+}$}
	\Text(160,15)[l]{$\longrightarrow$}
	\Text(185,17)[]{$\mu^{+}$}
	\Text(140,15)[r]{$\longleftarrow$}
	\Text(115,15)[]{$\nu_{\mu} $}
	\Text(167,22)[]{$\Leftarrow$}
	\Text(133,22)[]{$\Rightarrow$}
    \end{picture}   \end{center}
By angular momentum conservation, the spin projections of the muon 
and neutrino must sum to zero, so the helicity of the neutrino is 
equal to that of the muon: $h(\nu_{\mu}) = h(\mu^{+})$.  Note that 
because the massless neutrino must be left-handed, the $\mu^{+}$ is 
forced to have the ``wrong helicity'' in pion decay: the antilepton 
$\mu^{+}$ is naturally right-handed, and can only have a left-handed 
helicity because it is massive.  This ``helicity suppression'' 
inhibits the decay $\pi^{+} \to \mu^{+}\nu_{\mu}$, and it inhibits the 
analogue decay $\pi^{+} \to e^{+} \nu_{e}$ still more.  The decay 
amplitude in each case is proportional to the charged-lepton mass, 
and this accounts for the dramatic ratio
\begin{equation}
    \frac{\Gamma(\pi^{+} \to e^{+} \nu_{e})}
    {\Gamma(\pi^{+} \to \mu^{+}\nu_{\mu})} = 1.23 \times 10^{-4}\; .
    \label{eq:emurat}
\end{equation}
    
The classic determination of the electron neutrino's helicity was made
by M. Goldhaber and collaborators, who inferred $h(\nu_{e})$ from the
longitudinal polarization of the recoil nucleus in the
electron-capture reaction\cite{goldhaber}
\dknuc{e^{-} + ^{152}\mathrm{Eu}^{m} (J=0)}{^{152}\mathrm{Sm}^{*} 
(J=1)+\nu_{e}}{\gamma + ^{152}\mathrm{Sm}\; .}
Obviously, this achievement required not only impressive experimental 
technique, but also a remarkable knowledge of the characteristics of 
nuclear levels!

Recently a group in Z\"{u}rich has emulated the original achievement 
in a muon-capture reaction,
\begin{equation}
\mu^{-}\; ^{12}\mathrm{C} (J=0) \to \; ^{12}\mathrm{B} (J=1) \;\nu_{\mu}
\end{equation}
to determine $h(\nu_{\mu})$ by angular momentum
conservation.\cite{roesch} See the \textit{Review of Particle Physics}
for the most recent determinations of $h(\nu_{\tau})$ in tau
decays.\cite{pdg}

The fact that only left-handed neutrinos and right-handed 
antineutrinos are observed also means that charge-conjugation 
invariance is violated in the weak interactions.  Charge conjugation 
takes a $\nu_{L}$ into a nonexistent $\bar{\nu}_{L}$:
    \begin{center} \begin{picture}(300,30)(0,0)
	\Text(150,15)[]{$\mathcal{C}$}
	\Text(160,15)[l]{$\longrightarrow$  $\bar{\nu}_{L}$}
	\Text(182,15)[]{{\LARGE $/$}}
	\Text(182,15)[]{{\LARGE $\backslash$}}
	\Text(169,22)[]{$\Leftarrow$}
	\Text(140,15)[r]{$\nu_{L} \longrightarrow$}
	\Text(131,22)[]{$\Leftarrow \; .$}
    \end{picture}   \end{center}
The consequence of the $\mathcal{C}$ violation is very dramatic for muon 
decay, as the decay angular distributions of the outgoing $e^{\pm}$ in 
$\mu^{\pm}$ decay are reversed, 
\begin{equation}
    \frac{dN(\mu^{\pm}\to e^{\pm}+ \ldots)}{dxdz} =
    x^{2}(3-2x)\left[1 \pm z\frac{(2x-1)}{(3-2x)}\right]\; ,
    \label{eq:mudkang}
\end{equation}
where $x \equiv p_{e}/p_{e}^{\mathrm{max}}$ and $z \equiv 
\hat{s}_{\mu}\cdot\hat{p}_{e}$.  The positron follows the spin 
direction of the $\mu^{+}$, but the electron avoids the spin 
direction of the $\mu^{-}$.

\subsection{An effective Lagrangian for the weak interactions}
After the observation of maximal parity violation in the late 1950s, a 
serviceable effective Lagrangian for the weak interactions of 
electrons and neutrinos could be written as the product of charged 
leptonic currents,
\begin{equation}
    \lag_{V-A} = \frac{-G_{F}}{\sqrt{2}}\bar{\nu}\gamma_{\mu}
    (1 - \gamma_{5})e\; \bar{e}\gamma^{\mu}(1 - \gamma_{5})\nu
    + \mathrm{h.c.}\; ,
    \label{eq:lepel}
\end{equation}
where Fermi's coupling constant is $G_{F} = 1.16632 \times 
10^{-5}\gev^{-2}$.  This is often called the $V-A$ (vector minus axial 
vector) interaction. It is straightforward to compute the Feynman 
amplitude for antineutrino-electron scattering,\footnote{See 
Ref.\ \cite{gtswemi} for conventions and tricks for calculating 
amplitudes.} 
\begin{equation}
    \M = -\frac{iG_{F}}{\sqrt{2}} \bar{v}(\nu,q_{1})\gamma_{\mu} 
    (1 - \gamma_{5}) u(e,p_{1}) \bar{u}(e,p_{2}) \gamma^{\mu} (1 - 
    \gamma_{5}) v(\nu,q_{2})\; ,
    \label{eq:nubare}
\end{equation}
where the c.m.\ kinematical definitions are indicated in the sketch.
    \begin{center} \begin{picture}(100,110)(0,0)
	\ArrowLine(50,90)(50,50)
	\Text(50,95)[]{$e_{\mathrm{in}}: p_{1}$}
	\ArrowLine(50,10)(50,50)
	\Text(50,5)[]{$\nu_{\mathrm{in}}: q_{1}$}
	\ArrowLine(50,50)(78.28,78.28)
	\Text(80,80)[l]{$e_{\mathrm{out}}: p_{2}$}
	\ArrowLine(50,50)(21.72,21.72)
	\Text(20,20)[r]{$\nu_{\mathrm{out}}: q_{2}$}
	\CArc(50,50)(15,45,90)
	\Text(58.42,70.32)[l]{$\theta^{*}$}	
    \end{picture}   \end{center}
The differential cross section is related to the absolute square of 
the amplitude, averaged over initial spins and summed over final 
spins.  It is
\begin{equation}
    \frac{d\sigma_{V-A}(\bar{\nu}e \to \bar{\nu}e)}{d\Omega_{\mathrm{cm}}}  =  
    \frac{\overline{\abs{\M}^{2}}}{64\pi^{2}s}
    = \frac{G_{F}^{2}\cdot 2mE_{\nu}(1-z)^{2}}{16\pi^{2}}\; ,
    \label{eq:dsignubare}
\end{equation}
where $z = \cos\theta^{*}$.  The total cross section is simply
\begin{equation}
    \sigma_{V-A}(\bar{\nu}e \to \bar{\nu}e) = \frac{G_{F}^{2} \cdot 
    2mE_{\nu}}{3\pi} \approx 0.574 \times 10^{-41}\cm^{2}\left(
    \frac{E_{\nu}}{1\gev}\right) \; ;
    \label{eq:signubare}
\end{equation}
it is \textit{small!} for small energies.

Repeating the calculation for neutrino-electron scattering, we find
\begin{equation}
    \frac{d\sigma_{V-A}({\nu}e \to {\nu}e)}{d\Omega_{\mathrm{cm}}}  
    = \frac{G_{F}^{2}\cdot 2mE_{\nu}}{4\pi^{2}}\; ,
    \label{eq:dsignue}
\end{equation}
and
\begin{equation}
    \sigma_{V-A}({\nu}e \to {\nu}e) = \frac{G_{F}^{2} \cdot 
    2mE_{\nu}}{\pi} \approx 1.72 \times 10^{-41}\cm^{2}\left(
    \frac{E_{\nu}}{1\gev}\right) \; .
    \label{eq:signue}
\end{equation}
It is interesting to trace the origin of the factor-of-three 
difference between the $\nu e$ and $\bar{\nu}e$ cross sections, which 
arises from the left-handed nature of the charged current.  In 
neutrino-electron scattering, the initial state has spin projection 
$J_{z} = 0$, because the incoming neutrino and electron are both 
left -handed.  They can emerge in any direction---in particular, in the 
backward direction denoted by $z = +1$---and still satisfy the 
constraint that $J_{z} = 0$.
    \begin{center} \begin{picture}(280,70)(0,0)
	\ArrowLine(50,50)(50,30)
	\Text(50,55)[]{$e$}
	\Text(60,40)[]{$\Uparrow$}
	\ArrowLine(50,10)(50,30)
	\Text(50,5)[]{$\nu$}
	\Text(60,20)[]{$\Downarrow$}
	\Text(70,25)[l]{$J_{z}=0$}
	\Text(30,25)[r]{incoming}
	
	\ArrowLine(230,30)(230,50)
	\Text(230,55)[]{$e$}
	\Text(240,40)[]{$\Downarrow$}
	\ArrowLine(230,30)(230,10)
	\Text(230,5)[]{$\nu$}
	\Text(240,20)[]{$\Uparrow$}
	\Text(250,25)[l]{$J_{z}=0$}
	\Text(210,25)[r]{outgoing, $z=+1$}
	
    \end{picture}   \end{center}
In antineutrino-electron scattering, the situation is different, 
because the antineutrino is \textit{right-handed.}  The initial 
angular momentum has spin projection $J_{z} = 1$; for backward 
scattering, the outgoing electron and antineutrino combine to give 
$J_{z} = -1$, so scattering at $z=+1$ is forbidden by angular momentum 
conservation.
    \begin{center} \begin{picture}(280,70)(0,0)
	\ArrowLine(50,50)(50,30)
	\Text(50,55)[]{$e$}
	\Text(60,40)[]{$\Uparrow$}
	\ArrowLine(50,10)(50,30)
	\Text(50,5)[]{$\bar{\nu}$}
	\Text(60,20)[]{$\Uparrow$}
	\Text(70,25)[l]{$J_{z}=+1$}
	\Text(30,25)[r]{incoming}
	
	\ArrowLine(230,30)(230,50)
	\Text(230,55)[]{$e$}
	\Text(240,40)[]{$\Downarrow$}
	\ArrowLine(230,30)(230,10)
	\Text(230,5)[]{$\bar{\nu}$}
	\Text(240,20)[]{$\Downarrow$}
	\Text(250,25)[l]{$J_{z}=-1$}
	\Text(210,25)[r]{outgoing, $z=+1$}
	
    \end{picture}   \end{center}

\subsection{Lepton families and universality}
The muon is distinct from the electron; what is the nature of the 
neutrino emitted in pion decay, $\pi^{+} \to \mu^{+}\nu$?  In 1962, 
Lederman, Schwartz, Steinberger, and collaborators  carried out a 
\textit{two-neutrino experiment} using neutrinos created in the decay 
of high-energy pions from the new Alternating Gradient Synchrotron at 
Brookhaven.\cite{2nu}  They observed numerous examples of the 
reaction $\nu N \to \mu + X$, but found no evidence for the 
production of electrons.  Their study established that the muon 
produced in pion decay is a distinct particle, $\nu_{\mu}$, that is 
different from either $\nu_{e}$ or $\bar{\nu}_{e}$.  This observation 
suggests that the weak (charged-current) interactions of the leptons 
display a family structure,
\begin{equation}
 			\left(
		\begin{array}{c}
			\nu_{e}  \\
			e^{-}
		\end{array}
		 \right)_{L} \;\;\;\;\;
		\left(
		\begin{array}{c}
			\nu_{\mu}  \\
			\mu^{-}
		\end{array}
		 \right)_{L} \; .
    \label{eq:famlep}
\end{equation}
We are led to generalize the effective Lagrangian \eqn{eq:lepel} to 
include the terms
\begin{equation}
    \lag_{V-A}^{(e\mu)} = \frac{-G_{F}}{\sqrt{2}}\bar{\nu}_{\mu}\gamma_{\mu}
    (1 - \gamma_{5})\mu\; \bar{e}\gamma^{\mu}(1 - \gamma_{5})\nu_{e}
    + \mathrm{h.c.}\; ,
    \label{eq:lepmueel}
\end{equation}
in the familiar current-current form.  With this interaction, we 
easily compute the muon decay rate as
\begin{equation}
    \Gamma(\mu \to e \bar{\nu}_{e}\nu_{\mu}) = \frac{G_{F}^{2} 
        m_{\mu}^{5}}{192\pi^{3}}\; .
    \label{eq:mudk}
\end{equation}
With the value of the Fermi constant inferred from $\beta$ decay, 
\eqn{eq:mudk} accounts for the 2.2-$\mu$s lifetime of the muon.

The resulting cross section for inverse muon decay,
\begin{equation}
    \sigma(\nu_{\mu}e \to \mu \nu_{e}) = \sigma_{V-A}(\nu_{e}e \to 
    \nu_{e}e)\left[1 - \frac{(m_{\mu}^{2}- 
    m_{e}^{2})}{2m_{e}E_{\nu}}\right]^{2}\; ,
    \label{eq:invmudk}
\end{equation}
is in good agreement with high-energy data in measurements up to
$E_{\nu}\approx 600\gev$.  However, partial-wave unitarity constrains
the modulus of an inelastic amplitude to be $\abs{\M_{J}} < 1$. 
According to the $V-A$ theory, the $J=0$ partial-wave amplitude is
\begin{equation}
    \M_{0} = \frac{G_{F} \cdot 2m_{e}E_{\nu}}{\pi\sqrt{2}}
    \left[1 - \frac{(m_{\mu}^{2}-m_{e}^{2})}{2m_{e}E_{\nu}}\right] 
    \; ,
    \label{eq:invmudkpw}
\end{equation}
which satisfies the unitarity constraint for $E_{\nu} < 
\pi/G_{F}m_{e}\sqrt{2} \approx 3.7 \times 10^{8}\gev$.  These 
conditions aren't threatened anytime soon at an accelerator 
laboratory (though they do occur in interactions of cosmic 
neutrinos).  Nevertheless, we encounter here an important point of
principle: although the $V-A$ theory may be a reliable guide over a
broad range of energies, the theory cannot be complete: physics must
change before we reach a c.m.\ energy $\sqrt{s} \approx 600\gev$.

A few weeks after my TASI00 lectures, members of the \textsc{donut} (Direct Observation of 
NU Tau) experiment at Fermilab announced the first observation of 
charged-current interactions of the tau neutrino in a hybrid-emulsion 
detector situated in a ``prompt'' neutrino beam.\cite{donut} The 
$\nu_{\tau}$ beam was created in the production and decay of the 
charmed-strange meson
\dk{D_{s}^{+}}{\tau^{+}\nu_{\tau}}{\bar{\nu}_{\tau}+\hbox{anything.}} 
Their ``three-neutrino'' experiment was modeled on the two-neutrino 
classic: a beam of neutral, penetrating particles (the tau neutrinos) 
interacted in the hybrid target to produce tau leptons through the 
reaction 
\begin{equation}
    \nu_{\tau}N \to \tau + \hbox{anything.}
    \label{eq:tauapp}
\end{equation}
Although extensive studies of $\tau$ decays had given us a rather complete 
portrait of the interactions of $\nu_{\tau}$, the observation of the 
last of the known standard-model fermions gives a nice sense of 
closure, as well as a very impressive demonstration of the 
experimenter's art.

We have a great deal of precise information about the properties of 
the leptons, because the leptons are free particles readily studied 
in isolation.  All of them are spin-$\cfrac{1}{2}$, pointlike 
particles---down to a resolution of $\hbox{a few}\times 
10^{-17}\cm$.  The kinematically determined neutrino masses are all 
consistent with zero, though the evidence for neutrino oscillations 
argues that the neutrinos must have nonzero masses.  A brief digest 
of lepton properties is given in Table \ref{tbl:leptons}.
\begin{table}[tbp]
    \centering
    \caption{Some properties of the leptons.}
    \begin{tabular}{ccc}
        \hline
        Lepton & Mass & Lifetime  \\
        \hline\\[-6pt]
        $e^{-}$ & $0.510\,999\,07 \pm 0.000\,000\,15\mevcc$ & 
	$>4.3 \times 10^{23}\hbox{ y}\; (68\%\ \mathrm{CL})$  \\
        %\hline
        $\nu_{e}$ & $<10\hbox{ - }15\evcc$ &   \\[3pt]
        %\hline
        $\mu^{-}$ & $105.658\,389 \pm 0.000\,034\mevcc$ & $2.197\,03 
        \pm 0.000\,04\times 10^{-6}\s$  \\
        %\hline
        $\nu_{\mu}$ & $<0.19\mevcc\;(90\%\ \mathrm{CL})$ &   \\[3pt]
        %\hline
        $\tau^{-}$ & $1777.06^{+0.29}_{-0.26}\mevcc$ & $290.2 \pm 1.2 
        \times 10^{-15}\s$  \\
        %\hline
        $\nu_{\tau}$ & $<18.2\mevcc\;(95\%\ \mathrm{CL})$ &   \\[4pt]
        \hline
    \end{tabular}
    \label{tbl:leptons}
\end{table}

An important characteristic of the charged-current weak interactions 
is their \textit{universal strength,} which has been established in 
great detail.  We'll content ourselves here with the most obvious 
check for the lepton sector.  Using the generic formula \eqn{eq:mudk} for muon 
decay, we can use the measured lifetime of the muon to 
estimate the Fermi constant determined in muon decay as
\begin{equation}
    G_{\mu} = \left( \frac{192\pi^{3}\hbar}{\tau_{\mu}m_{\mu}^{5}}  
    \right)^{\cfrac{1}{2}} = 1.1638 \times 10^{-5}\gev^{-2}\; .
    \label{eq:Gmu}
\end{equation}
Similarly, we can evaluate the Fermi constant from the tau lifetime, 
taking into account the measured branching fraction for the leptonic 
decay.  We find
\begin{equation}
    G_{\tau} = \left( \frac{\Gamma(\tau \to e \bar{\nu}_{e}\nu_{\tau})}
    {\Gamma(\tau \to \hbox{all})} \cdot 
    \frac{192\pi^{3}\hbar}{\tau_{\tau}m_{\tau}^{5}}  
    \right)^{\cfrac{1}{2}} = 1.1642 \times 10^{-5}\gev^{-2}\; .
    \label{eq:Gtau}
\end{equation}
Both are in excellent agreement with the best value of the Fermi 
constant determined from nuclear $\beta$ decay,\footnote{In this 
discussion, but not in the number quoted, I'm glossing over the 
complication that the strangeness-preserving transition is not quite 
full (universal) strength.  We'll encounter ``Cabibbo universality'' 
in \S{\protect{\ref{sec:paradis}}}.} 
\begin{equation}
    G_{\beta} = 1.16639(2) \times 10^{-5}\gev^{-2}\; .
    \label{eq:Gbeta}
\end{equation}
The overall conclusion is that the charged currents acting in the 
leptonic and semileptonic interactions are of universal strength; we 
take this to imply a universality of the current-current form, or 
whatever lies behind it.

\section{The $\mathbf{SU(2)_{L}\otimes U(1)_{Y}}$ Electroweak Theory}
Let us review the essential elements of the \ws\ electroweak 
theory.\cite{gtswemi,GT,Gaillard:1998ui}
The electroweak theory takes three crucial clues from experiment:
\begin{itemize}
    \item  The existence of left-handed weak-isospin doublets,
    \begin{displaymath}
        \left( 
                \begin{array}{c}
            \nu_{e}  \\
            e
        \end{array}
	\right)_{L} \qquad
	        \left( 
                \begin{array}{c}
            \nu_{\mu}  \\
            \mu
        \end{array}
	\right)_{L} \qquad
        \left( 
                \begin{array}{c}
            \nu_{\tau}  \\
            \tau
        \end{array}
	\right)_{L}
    \end{displaymath}
    and
    \begin{displaymath}
        \left( 
                \begin{array}{c}
            u  \\
            d^{\prime}
        \end{array}
	\right)_{L} \qquad
	        \left( 
                \begin{array}{c}
            c  \\
            s^{\prime}
        \end{array}
	\right)_{L} \qquad
        \left( 
                \begin{array}{c}
            t  \\
            b^{\prime}
        \end{array}
	\right)_{L}\; ; 
    \end{displaymath}

    \item  The universal strength of the (charged-current) weak interactions;

    \item  The idealization that neutrinos are massless.
\end{itemize}

\subsection{A theory of leptons}
To save writing, we shall construct the electroweak theory as it 
applies to a single generation of leptons.  In this form, it is 
neither complete nor consistent: anomaly cancellation requires that a 
doublet of color-triplet quarks accompany each doublet of 
color-singlet leptons.  However, the needed generalizations are simple 
enough to make that we need not write them out.

To incorporate electromagnetism into a theory of the weak 
interactions, we add to the $SU(2)_{L}$ family symmetry suggested by 
the first two experimental clues a $U(1)_{Y}$ weak-hypercharge phase 
symmetry.\footnote{We define the weak hypercharge $Y$ through the 
Gell-Mann--Nishijima connection, $Q = I_{3} + \cfrac{1}{2}Y$, to 
electric charge and (weak) isospin.}
We begin by specifying the fermions: a left-handed weak 
isospin doublet
\begin{equation}
{{\sf L}} = \left(\begin{array}{c} \nu_e \\ e \end{array}\right)_L
\end{equation}
with weak hypercharge $Y_L=-1$, and a right-handed weak isospin singlet
\begin{equation}
      {{\sf R}}\equiv e_R
\end{equation}
with weak hypercharge $Y_R=-2$.

The electroweak gauge group, \ws, implies two sets of gauge fields:
a weak isovector $\vec{b}_\mu$, with coupling constant $g$, and a
weak isoscalar
${{\mathcal A}}_\mu$, with coupling constant $g^\prime$. Corresponding
to these gauge fields are the field-strength tensors 
\begin{equation}
    F^{\ell}_{\mu\nu} = \partial_{\nu}b^{\ell}_{\mu} - 
    \partial_{\mu}b^{\ell}_{\nu} + 
    g\varepsilon_{jk\ell}b^{j}_{\mu}b^{k}_{\nu}\; ,
    \label{eq:Fmunu}
\end{equation}
for the weak-isospin symmetry, and 
\begin{equation}
    f_{\mu\nu} = \partial_{\nu}{{\mathcal A}}_\mu - \partial_{\mu}{{\mathcal 
    A}}_\nu \; , 
    \label{eq:fmunu}
\end{equation}
for the weak-hypercharge symmetry.  We may summarize the interactions 
by the Lagrangian
\begin{equation}
\lag = \lag_{\rm gauge} + \lag_{\rm leptons} \ ,                           
\end{equation}             
with
\begin{equation}
\lag_{\rm gauge}=-\frac{1}{4}F_{\mu\nu}^\ell F^{\ell\mu\nu}
-\frac{1}{4}f_{\mu\nu}f^{\mu\nu},
\end{equation}
and
\begin{eqnarray}     
\lag_{\rm leptons} & = & \overline{{\sf R}}\:i\gamma^\mu\!\left(\partial_\mu
+i\frac{g^\prime}{2}{\cal A}_\mu Y\right)\!{\sf R} 
\label{eq:matiere} \\ 
& + & \overline{{\sf
L}}\:i\gamma^\mu\!\left(\partial_\mu 
+i\frac{g^\prime}{2}{\cal
A}_\mu Y+i\frac{g}{2}\vec{\tau}\cdot\vec{b}_\mu\right)\!{\sf L}. \nonumber
\end{eqnarray}
The \ws\ gauge symmetry forbids a mass term for the electron in the 
matter piece \eqn{eq:matiere}.  Moreover, the theory we have described 
contains four massless electroweak gauge bosons, namely ${{\mathcal A}}_\mu$, 
$b^{1}_{\mu}$, $b^{2}_{\mu}$, and $b^{3}_{\mu}$, whereas Nature has 
but one: the photon.  To give masses to the gauge bosons and 
constituent fermions, we must hide the electroweak symmetry.

The most apt analogy for the hiding of the electroweak gauge 
symmetry is found in superconductivity.  In the Ginzburg-Landau 
description\cite{4} of the superconducting phase transition, a 
superconducting material is regarded as a collection of two kinds of 
charge carriers: normal, resistive carriers, and superconducting, 
resistanceless carriers.

In the absence of a magnetic field, the free energy of the superconductor 
is related to the free energy in the normal state through
\begin{equation}
G_{\rm super}(0) = G_{\rm normal}(0) + \alpha \abs{\psi}^2 + \beta 
\abs{\psi}^4\;\;,
\end{equation}
where $\alpha$ and $\beta$ are phenomenological parameters and 
$\abs{\psi}^2$ is an order parameter that measures the density of 
superconducting charge carriers.  The parameter $\beta$ is non-negative, 
so that the free energy is bounded from below.

\begin{figure}
	\centerline{\BoxedEPSF{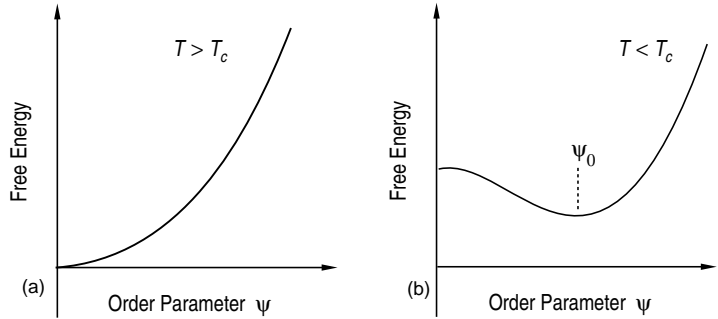  scaled 600}}
	\vspace*{6pt}
	\caption{Ginzburg-Landau description of the superconducting phase 
	transition.}
	\protect\label{fig1}
\end{figure}

Above the critical temperature for the onset of superconductivity, the 
parameter $\alpha$ is positive and the free energy of the substance is 
supposed to be an increasing function of the density of 
superconducting carriers, as shown in Figure \ref{fig1}(a).  The state 
of minimum energy, the vacuum state, then corresponds to a purely 
resistive flow, with no superconducting carriers active.  Below the 
critical temperature, the parameter $\alpha$ becomes negative and the 
free energy is minimized when $\psi = \psi_0 = \sqrt{-\alpha/\beta} \ne 0$, as illustrated in 
Figure \ref{fig1}(b).

This is a nice cartoon description of the superconducting phase 
transition, but there is more.  In an applied magnetic field $\vec{H}$, 
the free energy is
\begin{equation}
G_{\rm super}(\vec{H}) = G_{\rm super}(0) + \frac{\vec{H}^2}{8\pi} + 
\frac{1}{2m^\star}|-i\hbar\nabla\psi-(e^\star/c)\vec{A}\psi|^2 
\;\;,
\end{equation}
where $e^\star$ and $m^\star$ are the charge ($-2$ units) and effective 
mass of the superconducting carriers.  In a weak, slowly varying field 
$\vec{H} \approx 0$, when we can approximate $\psi\approx\psi_0$ and 
$\nabla\psi\approx 0$, the usual variational analysis leads to the 
equation of motion,
\begin{equation}
\nabla^2\vec{A}-\frac{4\pi e^\star}{m^\star c^2}\abs{\psi_0}^2\vec{A} = 
0\;\;,
\end{equation}
the wave equation of a massive photon.  In other words, the photon 
acquires a mass within the superconductor.  This is the origin of the 
Meissner effect, the exclusion of a magnetic field from a 
superconductor.  More to the point for our purposes, it shows how a 
symmetry-hiding phase transition can lead to a massive gauge boson.

To give masses to the intermediate bosons of the weak interaction, we 
take advantage of a relativistic generalization of the Ginzburg-Landau 
phase transition known as the Higgs mechanism.\cite{5}  We introduce 
a complex doublet of scalar fields
\begin{equation}
\phi\equiv \left(\begin{array}{c} \phi^+ \\ \phi^0 \end{array}\right)
\end{equation}
with weak hypercharge $Y_\phi=+1$.  Next, we add to the Lagrangian new 
(gauge-invariant) terms for the interaction and propagation of the 
scalars,
\begin{equation}
      \lag_{\rm scalar} = (\D^\mu\phi)^\dagger(\D_\mu\phi) - V(\phi^\dagger \phi),
\end{equation}
where the gauge-covariant derivative is
\begin{equation}
      \D_\mu=\partial_\mu 
+i\frac{g^\prime}{2}{\cal A}_\mu
Y+i\frac{g}{2}\vec{\tau}\cdot\vec{b}_\mu \; ,
\label{eq:GcD}
\end{equation}
and the potential interaction has the form
\begin{equation}
      V(\phi^\dagger \phi) = \mu^2(\phi^\dagger \phi) +
\abs{\lambda}(\phi^\dagger \phi)^2 .
\label{SSBpot}
\end{equation}
We are also free to add a Yukawa interaction between the scalar fields
and the leptons,
\begin{equation}
      \lag_{\rm Yukawa} = -\zeta_e\left[\overline{{\sf R}}(\phi^\dagger{\sf
L}) + (\overline{{\sf L}}\phi){\sf R}\right].
\label{eq:Yukterm}
\end{equation}

We then arrange 
their self-interactions so that the vacuum state corresponds to a 
broken-symmetry solution.  The electroweak symmetry is spontaneously broken if the parameter
$\mu^2<0$. The minimum energy, or vacuum state, may then be chosen
to correspond to the vacuum expectation value
\begin{equation}
\vev{\phi} = \left(\begin{array}{c} 0 \\ v/\sqrt{2} \end{array}
\right),
\label{eq:vevis}
\end{equation}
where $v = \sqrt{-\mu^2/\abs{\lambda}}$.

Let us verify that the vacuum \eqn{eq:vevis} indeed breaks the gauge 
symmetry.  The vacuum state $\vev{\phi}$ is invariant under a symmetry 
operation $\exp{(i \alpha {\mathcal G})}$ corresponding to the 
generator ${\mathcal G}$ provided that $\exp{(i \alpha {\mathcal 
G})}\vev{\phi} = \vev{\phi}$, \ie, if ${\mathcal G}\vev{\phi} = 0$.  
We easily compute that 
\begin{eqnarray}
    \tau_{1}\vev{\phi} & = \left( 
    \begin{array}{cc}
        0 & 1  \\
        1 & 0
    \end{array}
    \right) \left( 
    \begin{array}{c}
        0  \\
        v/\sqrt{2}
    \end{array}
    \right) & = \left( 
    \begin{array}{c}
        v/\sqrt{2}  \\
        0
    \end{array}
    \right) \neq 0 \quad\hbox{broken!}
    \nonumber  \\
    \tau_{2}\vev{\phi} & = \left( 
    \begin{array}{cc}
        0 & -i  \\
        i & 0
    \end{array}
    \right) \left( 
    \begin{array}{c}
        0  \\
        v/\sqrt{2}
    \end{array}
    \right) & = \left( 
    \begin{array}{c}
        -iv/\sqrt{2}  \\
        0
    \end{array}
    \right) \neq 0 \quad\hbox{broken!}
    \nonumber  \\
    \tau_{3}\vev{\phi} & = \left( 
    \begin{array}{cc}
        1 & 0  \\
        0 & -1
    \end{array}
    \right) \left( 
    \begin{array}{c}
        0  \\
        v/\sqrt{2}
    \end{array}
    \right) & = \left( 
    \begin{array}{c}
        0  \\
        -v/\sqrt{2}
    \end{array}
    \right) \neq 0 \quad\hbox{broken!}
    \nonumber  \\
    Y\vev{\phi} & = Y_{\phi}\vev{\phi} = +1 \vev{\phi} = & \left( 
    \begin{array}{c}
        0  \\
        v/\sqrt{2}
    \end{array}
    \right) \neq 0 \quad\hbox{broken!}
    \label{eq:brisure}
\end{eqnarray}
However, if we examine the effect of the electric charge operator $Q$ 
on the (electrically neutral) vacuum state, we find that
\begin{eqnarray}
    Q \vev{\phi} & = \cfrac{1}{2}(\tau_{3} + Y)\vev{\phi} & = 
    \cfrac{1}{2} \left( 
    \begin{array}{cc}
        Y_{\phi}+1 & 0  \\
        0 & Y_{\phi}- 1
    \end{array}
    \right) \vev{\phi}
    \nonumber  \\
     & = \left( 
     \begin{array}{cc}
         1 & 0  \\
         0 & 0
     \end{array}
      \right) \left(     \begin{array}{c}
        0  \\
        v/\sqrt{2}
    \end{array}
\right) & = \left( 
     \begin{array}{c}
         0  \\
         0
     \end{array}
     \right) \quad\hbox{\textit{unbroken!}}
    \label{eq:Qok}
\end{eqnarray}
The original four generators are all broken, but electric charge is
not.  It appears that we have accomplished our goal of breaking
$SU(2)_{L}\otimes U(1)_{Y} \to U(1)_{\mathrm{em}}$.  We expect the
photon to remain massless, and expect the gauge bosons that correspond
to the generators $\tau_{1}$, $\tau_{2}$, and $\kappa \equiv
\cfrac{1}{2}(\tau_{3} - Y)$ to acquire masses.

As a result of spontaneous symmetry breaking, the weak bosons acquire 
masses, as auxiliary scalars assume the role of the third 
(longitudinal) degrees of freedom of what had been massless gauge 
bosons.  Specifically, the mediator of the charged-current weak 
interaction, $W^{\pm} = (b_{1} \mp ib_{2})/\sqrt{2}$, acquires a 
mass characterized by 
\begin{equation}
    M_{W} = \frac{gv}{2} \; .
    \label{eq:wmassdef}
\end{equation}
With the definition $g^{\prime} = g\tan\theta_{W}$, where $\theta_{W}$ 
is the weak mixing angle, the mediator of the neutral-current weak 
interaction, $Z = b_{3}\cos{\theta_{W}} - \mathcal{A}\sin{\theta_{W}}$, 
acquires a mass characterized by 
$M_Z^2=M_W^2/\cos^2{\theta_W}$.  After spontaneous symmetry breaking, 
there remains an unbroken $U(1)_{\mathrm{em}}$ phase symmetry, so that 
electromagnetism, a vector interaction, is mediated by a massless photon, $A = 
\mathcal{A}\cos{\theta_{W}} + b_{3}\sin{\theta_{W}}$, coupled to the 
electric charge $e = gg^{\prime}/\sqrt{g^{2} + g^{\prime 2}}$.  As a vestige 
of the spontaneous breaking of the symmetry, there remains a massive, 
spin-zero particle, the Higgs boson.  The mass of the Higgs scalar is 
given symbolically as $M_{H}^{2} = -2\mu^{2} > 0$, but we have no 
prediction for its value.  Though what we take to be the work of the 
Higgs boson is all around us, the Higgs particle itself has not yet 
been observed.

The fermions (the electron in our abbreviated treatment) acquire 
masses as well; these are determined not only by the scale of 
electroweak symmetry breaking, $v$, but also by  their Yukawa interactions with
the scalars.  The mass of the electron is set by the dimensionless 
coupling constant $\zeta_{e} = m_{e}\sqrt{2}/v$, 
which is---so far as we now know---arbitrary.

\subsection{The $W$ boson {\protect{\label{sec:IVB}}}} 
The interactions of the $W$-boson with the leptons are given by 
\begin{equation}
    \lag_{W\mathrm{-lep}} = \frac{-g}{2\sqrt{2}}\left[ 
    \bar{\nu}_{e}\gamma^{\mu}(1 - \gamma_{5})eW^{+}_{\mu} +
    \bar{e}\gamma^{\mu}(1 - \gamma_{5})\nu_{e}W^{-}_{\mu}
    \right]\; , \mathrm{etc.},
    \label{eq:Wleplag}
\end{equation}
so the Feynman rule for the $\nu_{e}eW$ vertex is
\begin{center} \begin{picture}(160,100)(0,0)
	\ArrowLine(30,50)(10,90)
	\Text(5,85)[]{$e$}
	\ArrowLine(10,10)(30,50)
	\Text(5,15)[]{$\nu$}
	\ZigZag(30,50)(90,50){3}{6}
	\Text(95,58)[]{$\lambda$}
	\Text(110,50)[l]{${\displaystyle\frac{-ig}{2\sqrt{2}}}\gamma_{\lambda}(1 - \gamma_{5})$}
\end{picture}   \end{center}
The $W$-boson propagator is     \begin{picture}(100,15)(0,0)
        \ZigZag(0,0)(60,0){3}{4}
        \Text(75,0)[l]{$= \displaystyle{\frac{-i(g_{\mu\nu} - 
                k_{\mu}k_{\nu}/M_{W}^{2})}{k^{2} - M_{W}^{2}}}\; .$}
    \end{picture}\\[12pt]
%\vspace*{6pt}    
Let us compute the cross section for inverse muon decay in the new 
theory.   We find
\begin{equation}
    \sigma(\nu_{\mu}e \to \mu\nu_{e}) = 
    \frac{g^{4}m_{e}E_{\nu}}{16\pi M_{W}^{4}} \; \frac{\left[ 1 - 
    (m_{\mu}^{2} - m_{e}^{2})/2m_{e}E_{\nu}\right]^{2}}{(1 + 
    2m_{e}E_{\nu}/M_{W}^{2})}\; ,
    \label{eq:invmudkW}
\end{equation}
which coincides with the four-fermion result \eqn{eq:invmudk} at low 
energies, provided we identify
\begin{equation}
    \frac{g^{4}}{16M_{W}^{2}} = 2 G_{F}^{2}\; ,
    \label{eq:gidGF}
\end{equation}
which implies that
\begin{equation}
    \frac{g}{2\sqrt{2}} = 
    \left(\frac{G_{F}M_{W}^{2}}{\sqrt{2}}\right)^{\cfrac{1}{2}}\; .
    \label{eq:gidGF3}
\end{equation}
With the aid of \eqn{eq:wmassdef} for the $W$-boson mass, we 
determine the numerical value,
\begin{equation}
    v = \left(G_{F}\sqrt{2}\right)^{-\cfrac{1}{2}} \approx 246\gev\; .
    \label{eq:vevval}
\end{equation}

The high-energy limit of the cross section \eqn{eq:invmudkW} is
\begin{equation}
    \lim_{E_{\nu}\to \infty}\sigma(\nu_{\mu}e \to \mu\nu_{e}) = 
    \frac{g^{4}}{32\pi M_{W}^{2}} = \frac{G_{F}M_{W}^{2}}{\pi} \;,
    \label{eq:invmudklim}
\end{equation}
independent of energy.  The benign high-energy behavior means that 
partial-wave unitarity is now respected for 
\begin{equation}
    s < M_{W}^{2}\left[ \exp\left(\frac{\pi\sqrt{2}}{G_{F}M_{W}^{2}} \right) - 
    1 \right]\; ,
    \label{eq:pwuW}
\end{equation}
an immense improvement over the four-fermion theory.

Let us now investigate 
the properties of the $W$-boson in terms of its mass, $M_{W}$.  
Consider first the leptonic disintegration of the $W^{-}$, with decay 
kinematics specified thus:
    \begin{center} \begin{picture}(200,100)(0,0)
	\ArrowLine(20,50)(20,90)
	\Text(28,80)[l]{$e(p)$\qquad$\displaystyle{p\approx\left(\frac{M_{W}}{2};
	    \frac{M_{W}\sin\theta}{2}, 0, \frac{M_{W}\cos\theta}{2}\right)}$}
	\ArrowLine(20,50)(20,10)
	\Text(28,20)[l]{$\bar{\nu}_{e}(q)$\qquad$\displaystyle{q\approx\left(\frac{M_{W}}{2};
	    -\,\frac{M_{W}\sin\theta}{2}, 0, -\,\frac{M_{W}\cos\theta}{2}\right)}$}
	\Vertex(20,50){3}
	\Text(12,50)[r]{$W^{-}$}
    \end{picture}   \end{center}
The Feynman amplitude for the decay is
\begin{equation}
    \M = -i \left( \frac{G_{F}M_{W}^{2}}{\sqrt{2}}\right)^{\cfrac{1}{2}}
    \bar{u}(e,p)\gamma_{\mu}(1 - \gamma_{5})v(\nu,q)\, 
    \varepsilon^{\mu}\; ,
    \label{eq:Wdkamp}
\end{equation}
where $\varepsilon^{\mu}= (0; \hat{\varepsilon})$ is the polarization 
vector of the $W$-boson in its rest frame.  The square of the 
amplitude is
\begin{eqnarray}
    \abs{\M}^{2} & = & \frac{G_{F}M_{W}^{2}}{\sqrt{2}}
    \tr{\left[ \slashi{\varepsilon}(1-\gamma_{5})\slashiv{q}(1+\gamma_{5})
    \slashi{\varepsilon}^{*}\slashiv{p}\right]}  
    \label{eq:Wdk2} \\
     & = & \displaystyle{\frac{8G_{F}M_{W}^{2}}{\sqrt{2}}}\left[ 
     \varepsilon\cdot q \: \varepsilon^{*}\cdot p -
     \varepsilon \cdot \varepsilon^{*} \: q\cdot p +
     \varepsilon \cdot p \: \varepsilon^{*}\cdot q +
     i\epsilon_{\mu\nu\rho\sigma}\varepsilon^{\mu}q^{\nu}\varepsilon^{*\rho}p^{\sigma}
     \right] \nonumber \; .
\end{eqnarray}
The \textit{decay rate} is independent of the $W$ polarization, so 
let us look first at the case of longitudinal polarization 
$\varepsilon^{\mu}=(0;0,0,1)=\varepsilon^{*\mu}$, to eliminate the 
last term.  For this case, we find
\begin{equation}
     \abs{\M}^{2} = \frac{4G_{F}M_{W}^{4}}{\sqrt{2}}\sin^{2}\theta\; ,
    \label{eq:Wdk3}
\end{equation}
so the differential decay rate is
\begin{equation}
    \frac{d\Gamma_{0}}{d\Omega} =  \frac{\abs{\M}^{2}}{64\pi^{2}} \: 
    \frac{{\mathcal S}_{12}}{M_{W}^{3}} \; ,
    \label{eq:Wdk4}
\end{equation}
where ${\mathcal S}_{12} = \sqrt{[M_{W}^{2}-(m_{e}+m_{\nu})^{2}]
[M_{W}^{2}-(m_{e}-m_{\nu})^{2}]} = M_{W}^{2}$, so that
\begin{equation}
    \frac{d\Gamma_{0}}{d\Omega} = 
    \frac{G_{F}M_{W}^{3}}{16\pi^{2}\sqrt{2}}\sin^{2}\theta \; ,
    \label{eq:Wdk5}
\end{equation}
and 
\begin{equation}
    \Gamma(W \to e\nu) = \frac{G_{F}M_{W}^{3}}{6\pi\sqrt{2}} \; .
    \label{eq:Wdktot}
\end{equation}

For the other helicities, $\varepsilon^{\mu}_{\pm 1} = (0; -1, \mp i, 
0)/\sqrt{2}$, arithmetic that is only slightly more 
tedious leads us to 
\begin{equation}
    \frac{d\Gamma_{\pm 1}}{d\Omega} = 
    \frac{G_{F}M_{W}^{3}}{32\pi^{2}\sqrt{2}}(1 \mp \cos\theta)^{2} \;   .  
    \label{eq:Wdkpm1}
\end{equation}
The extinctions at $\cos\theta = \pm 1$ are, as we have come to 
expect, consequences of angular momentum conservation:
    \begin{center} \begin{picture}(280,70)(0,0)
	\Text(0,35)[l]{$W^{-}$\quad{\Huge$ \Uparrow$}}
	\ArrowLine(70,30)(70,50)
	\Text(70,55)[]{$e^{-}$}
	\Text(80,40)[]{$\Downarrow$}
	\ArrowLine(70,30)(70,10)
	\Text(70,5)[]{$\bar{\nu}_{e}$}
	\Text(80,20)[]{$\Downarrow$}
	\Text(90,25)[l]{$(\theta=0)$ \textit{forbidden}}
	\ArrowLine(190,30)(190,50)
	\Text(190,55)[]{$\bar{\nu}_{e}$}
	\Text(200,40)[]{$\Uparrow$}
	\ArrowLine(190,30)(190,10)
	\Text(190,5)[]{$e^{-}$}
	\Text(200,20)[]{$\Uparrow$}
	\Text(210,25)[l]{$(\theta=\pi)$ allowed}
    \end{picture}   \end{center}
The situation is reversed for the decay of $W^{+} \to e^{+}\nu_{e}$.  
Overall, the $e^{+}$ follows the polarization direction of $W^{+}$, 
while the $e^{-}$ avoids the polarization direction of $W^{-}$.  This 
charge asymmetry was important for establishing the discovery of the 
$W$-boson in $\bar{p}p$ ($\bar{q}q$) collisions.

\subsection{Neutral Currents}
The interactions of the $Z$-boson with leptons are given by 
\begin{equation}
    \lag_{Z\mathrm{-}\nu} = \frac{-g}{4\cos\theta_{W}} 
    \bar{\nu}\gamma^{\mu}(1 - \gamma_{5})\nu\:Z_{\mu}
    \label{eq:Znulag}
\end{equation}
and
\begin{equation}
    \lag_{Z\mathrm{-}e} = \frac{-g}{4\cos\theta_{W}} 
    \bar{e}\left[L_{e}\gamma^{\mu}(1 - \gamma_{5}) +
    R_{e}\gamma^{\mu}(1 + \gamma_{5})\right]e\:Z_{\mu}\; ,
    \label{eq:Zelag}
\end{equation}
where the chiral couplings are
\begin{eqnarray}
    L_{e} & = & 2 \sin^{2}\theta_{W} - 1 = 2x_{W} + \tau_{3} \; ,
    \nonumber  \\
    R_{e} & = & 2 \sin^{2}\theta_{W}\; .
    \label{eq:lepchicoup}
\end{eqnarray}
By analogy with the calculation of the $W$-boson total width 
\eqn{eq:Wdktot}, we easily compute that
\begin{eqnarray}
    \Gamma(Z \to \nu\bar{\nu}) & = & \frac{G_{F}M_{Z}^{3}}{12\pi\sqrt{2}}
    \nonumber \\
    \Gamma(Z \to e^{+}e^{-}) & = & \Gamma(Z \to 
    \nu\bar{\nu})\left[L_{e}^{2} + R_{e}^{2}\right]\; .
    \label{eq:Zwidths}
\end{eqnarray}

The neutral weak current mediates a reaction that did not arise in 
the $V-A$ theory, $\nu_{\mu}e \to \nu_{\mu}e$, which proceeds 
entirely by $Z$-boson exchange:
\begin{center} \begin{picture}(130,100)(0,0)
	\ArrowLine(30,50)(10,90)
	\Text(5,85)[]{$\nu_{\mu}$}
	\ArrowLine(10,10)(30,50)
	\Text(5,15)[]{$\nu_{\mu}$}
	\ZigZag(30,50)(100,50){4}{5}
	\ArrowLine(100,50)(120,90)
	\Text(125,85)[]{$e$}
	\ArrowLine(120,10)(100,50)
	\Text(125,5)[]{$e$}
   \end{picture}   \end{center}
This was, in fact, the reaction in which the first evidence for the 
weak neutral current was seen by the Gargamelle collaboration in 
1973.\cite{GGM}   It's an easy exercise to compute all the cross 
sections for neutrino-electron elastic scattering.  We find
\begin{eqnarray}
    \sigma(\nu_{\mu}e \to \nu_{\mu}e) & = &
    \displaystyle{\frac{G_{F}^{2}m_{e}E_{\nu}}{2\pi}} \left[L_{e}^{2} + 
    R_{e}^{2}/3\right]  \; ,
    \nonumber  \\
    \sigma(\bar{\nu}_{\mu}e \to \bar{\nu}_{\mu}e) & = & 
    \displaystyle{\frac{G_{F}^{2}m_{e}E_{\nu}}{2\pi}} \left[L_{e}^{2}/3 + 
    R_{e}^{2}\right]  \; ,
    \nonumber  \\
    \sigma(\nu_{e}e \to \nu_{e}e) & = &
    \displaystyle{\frac{G_{F}^{2}m_{e}E_{\nu}}{2\pi}} \left[(L_{e}+2)^{2} + 
    R_{e}^{2}/3\right]  \; ,
    \nonumber  \\
    \sigma(\bar{\nu}_{e}e \to \bar{\nu}_{e}e) & = & 
    \displaystyle{\frac{G_{F}^{2}m_{e}E_{\nu}}{2\pi}} \left[(L_{e}+2)^{2}/3 + 
    R_{e}^{2}\right]  \; .
    \label{eq:signueel}
\end{eqnarray}
By measuring all the cross sections, one may undertake a 
``model-independent'' determination\footnote{It is model-independent 
within the framework of vector and axial-vector couplings only, so in 
the context of gauge theories.} of the chiral couplings $L_{e}$ 
and $R_{e}$, or the traditional vector and axial-vector couplings $v$ 
and $a$, which are related through
\begin{equation}
    \begin{array}{ccc}
        a = \cfrac{1}{2}(L_{e}-R_{e}) & \quad & v = 
        \cfrac{1}{2}(L_{e}-R_{e})  \\[3pt]
        L_{e} = v+a & \quad & R_{e} = v-a
    \end{array}\; .
    \label{eq:chiva}
\end{equation}
By inspecting \eqn{eq:signueel}, you can see that even after measuring 
all four cross sections, there remains a two-fold ambiguity: the same 
cross sections result if we interchange $R_{e} \leftrightarrow 
-R_{e}$, or, equivalently, $v \leftrightarrow a$.  The ambiguity is 
resolved by measuring the forward-backward asymmetry in a reaction 
like $e^{+}e^{-} \to \mu^{+}\mu^{-}$ at energies well below the 
$Z^{0}$ mass.  The asymmetry is proportional to 
$(L_{e}-R_{e})(L_{\mu}-R_{\mu})$, or to $a_{e}a_{\mu}$, and so 
resolves the sign ambiguity for $R$ or the $v$-$a$ ambiguity.
\subsection{Electroweak interactions of quarks}
To extend our theory to include the electroweak interactions of 
quarks, we observe that each generation consists of a left-handed 
doublet
\begin{equation}
    \begin{array}{cccc}
         & I_{3} & Q & Y = 2(Q - I_{3})  \\[6pt]
        \mathsf{L}_{q}= \left( 
        \begin{array}{c}
            u  \\[3pt]
            d
        \end{array}
        \right)_{L}\quad & 
        \begin{array}{c}
            \cfrac{1}{2}  \\[3pt]
            - \cfrac{1}{2}
        \end{array}
         & 
         \begin{array}{c}
             +\cfrac{2}{3}  \\[3pt]
             -\cfrac{1}{3}
         \end{array}
          & \cfrac{1}{3}\; ,
    \end{array}
    \label{eq:LHq}
\end{equation}
and two right-handed singlets, 
\begin{equation}
    \begin{array}{cccc}
         & I_{3} & Q & Y = 2(Q - I_{3})  \\[6pt]
         
        \begin{array}{c}
            \mathsf{R}_{u} = u_{R}  \\[3pt]
            \mathsf{R}_{d} = d_{R}
        \end{array}
        \quad & 
        \begin{array}{c}
            0  \\[3pt]
            0
        \end{array}
         & 
         \begin{array}{c}
             +\cfrac{2}{3}  \\[3pt]
             -\cfrac{1}{3}
         \end{array}
          &          \begin{array}{c}
             +\cfrac{4}{3}  \\[3pt]
             -\cfrac{2}{3}
         \end{array}\; ,

    \end{array}
    \label{eq:RHq}
\end{equation}
Proceeding as before, we find the Lagrangian terms for the $W$-quark 
charged-current interaction,
\begin{equation}
    \lag_{W\mathrm{-quark}} = \frac{-g}{2\sqrt{2}}\left[ 
    \bar{u}_{e}\gamma^{\mu}(1 - \gamma_{5})d \:W^{+}_{\mu} +
    \bar{d}\gamma^{\mu}(1 - \gamma_{5})u \:W^{-}_{\mu}
    \right]\; , 
    \label{eq:Wqlag}
\end{equation}
which is identical in form to the leptonic charged-current 
interaction \eqn{eq:Wleplag}.  Universality is ensured by the fact 
that the charged-current interaction is determined by the weak 
isospin of the fermions, and that both quarks and leptons come in 
doublets.

The neutral-current interaction is also equivalent in form to its 
leptonic counterpart, \eqn{eq:Znulag} and \eqn{eq:Zelag}.  We may write it compactly as
\begin{equation}
    \lag_{Z\mathrm{-quark}} = \frac{-g}{4\cos\theta_{W}} 
    \sum_{i=u,d} \bar{q}_{i}\gamma^{\mu}\left[ L_{i}(1 - 
    \gamma_{5}) + R_{i}(1 + \gamma_{5})\right]q_{i}\:Z_{\mu}\; ,
    \label{eq:Zqlag}
\end{equation}
where the chiral couplings are
\begin{eqnarray}
    L_{i} & = & \tau_{3} - 2Q_{i} \sin^{2}\theta_{W} \; ,
    \nonumber  \\
    R_{i} & = & -2Q_{i} \sin^{2}\theta_{W}\; .
    \label{eq:qchicoup}
\end{eqnarray}
Again we find a quark-lepton universality in the form---but not the 
values---of the chiral couplings.

\subsection{Trouble in Paradise \label{sec:paradis}}
Until now, we have based our construction on the idealization that 
the $u \leftrightarrow d$ transition is of universal strength.  The 
unmixed doublet
\begin{displaymath}
    \left( 
    \begin{array}{c}
        u  \\
        d
    \end{array}
    \right)_{L}
\end{displaymath}
does not quite describe our world.  We attain a better description by 
replacing
\begin{displaymath}
    \left( 
    \begin{array}{c}
        u  \\
        d
    \end{array}
    \right)_{L} \to
    \left( 
    \begin{array}{c}
        u  \\
        d_{\theta}
    \end{array}
    \right)_{L} \; ,   
\end{displaymath}
where 
\begin{equation}
    d_{\theta} \equiv d\,\cos\theta_{C} + s\,\sin\theta_{C}\; ,
    \label{eq:cabrot}
\end{equation}
with $\cos\theta_{C} = 0.9736 \pm 0.0010$.\footnote{The arbitrary 
Yukawa couplings that give masses to the quarks can easily be chosen 
to yield this result.}  The change to the ``Cabibbo-rotated'' doublet 
perfects the charged-current interaction---at least up to small 
third-generation effects that we could easily incorporate---but leads 
to serious trouble in the neutral-current sector, for which the 
interaction now becomes
\begin{eqnarray}
    \lag_{Z\mathrm{-quark}} & = & \frac{-g}{4\cos\theta_{W}} \:Z_{\mu}
     \left\{\bar{u}\gamma^{\mu}\left[ L_{u}(1 - 
    \gamma_{5}) + R_{u}(1 + \gamma_{5})\right]u \right. \nonumber \\
      & & + \bar{d}\gamma^{\mu}\left[ L_{d}(1 - 
    \gamma_{5}) + R_{d}(1 + 
    \gamma_{5})\right]d\,\cos^{2}\theta_{C}  \nonumber \\
      & &   + \bar{s}\gamma^{\mu}\left[ L_{d}(1 - 
    \gamma_{5}) + R_{d}(1 + \gamma_{5})\right]s\,\sin^{2}\theta_{C} 
    \nonumber \\
      & &  +  \bar{d}\gamma^{\mu}\left[ L_{d}(1 - 
    \gamma_{5}) + R_{d}(1 + 
    \gamma_{5})\right]s\,\sin\theta_{C}\cos\theta_{C} 
    \nonumber \\
      & &   + \left. \bar{s}\gamma^{\mu}\left[ L_{d}(1 - 
    \gamma_{5}) + R_{d}(1 + 
    \gamma_{5})\right]d\,\sin\theta_{C}\cos\theta_{C} \right\}
    \; ,
    \label{eq:Zqrotlag}
\end{eqnarray}
Until the discovery and systematic study of the weak neutral current, 
culminating in the heroic measurements made at LEP and the SLC, there 
was not enough knowledge to challenge the first three terms.  The last 
two \textit{strangeness-changing} terms were known to be poisonous, 
because many of the early experimental searches for neutral currents 
were fruitless searches for precisely this sort of interaction.  
Strangeness-changing neutral-current interactions are not seen at an 
impressively low level.\footnote{For more on rare kaon decays, see 
the TASI 2000 lectures by Tony Barker\cite{TonyB} and Gerhard 
Buchalla.\cite{GerhardB}}

Only recently has Brookhaven Experiment 787\cite{E787} detected a 
single candidate for the decay $K^{+} \to \pi^{+}\nu\bar{\nu}$,
    \begin{center} \begin{picture}(100,80)(0,0)
	\ArrowLine(50,30)(20,30)
	\ArrowLine(80,30)(50,30)
	\Vertex(50,30){2}
	\ArrowLine(20,20)(80,20)
	\Text(17,25)[r]{$K^{+}$}
	\Text(85,25)[bl]{$\pi^{+}$}
	\Text(25,35)[b]{$\bar{s}$}
	\Text(75,35)[b]{$\bar{d}$}
	\Text(25,15)[t]{$u$}
	\ZigZag(50,30)(65,55){2}{5}
	\ArrowLine(60,70)(65,55)
	\Text(60,76)[]{$\bar{\nu}$}
	\Text(85,65)[]{$\nu$}
	\ArrowLine(65,55)(80,60)
    \end{picture}   \end{center}
and inferred a branching ratio ${\mathcal B}(K^{+} \to 
\pi^{+}\nu\bar{\nu}) = 1.5^{+3.5}_{-1.3}\times 10^{-10}$.  The good 
agreement between the standard-model prediction, ${\mathcal B}(K_{L} 
\to \mu^{+}\mu^{-}) = 0.8 \pm 0.3 \times 10^{-10}$ (through the 
process $K_{L} \to \gamma\gamma \to \mu^{+}\mu^{-}$), and 
experiment\cite{Kmumu} 
leaves little room for a strangeness-changing neutral-current 
contribution:  
    \begin{center} \begin{picture}(150,50)(0,0)
	\ArrowLine(70,30)(20,30)
	\ArrowLine(20,20)(70,20)
	\CArc(70,25)(5,-90,90)
	\Vertex(75,25){2}
	\ZigZag(75,25)(115,25){2}{5}
	\ArrowLine(125,50)(115,25)
	\ArrowLine(115,25)(125,0)
	
	\Text(17,25)[r]{$K^{+}$}
	\Text(130,45)[l]{$\mu^{+}$}
	\Text(130,5)[l]{$\mu^{-}$}
	\Text(25,35)[b]{$\bar{s}$}
	\Text(25,15)[t]{$d$}
	\Text(149,25)[r]{,}
    \end{picture}   \end{center}
that is easily normalized to the normal charged-current leptonic decay 
of the $K^{+}$:
    \begin{center} \begin{picture}(150,50)(0,0)
	\ArrowLine(70,30)(20,30)
	\ArrowLine(20,20)(70,20)
	\CArc(70,25)(5,-90,90)
	\Vertex(75,25){2}
	\ZigZag(75,25)(115,25){2}{5}
	\ArrowLine(125,50)(115,25)
	\ArrowLine(115,25)(125,0)
	
	\Text(17,25)[r]{$K^{+}$}
	\Text(130,45)[l]{$\mu^{+}$}
	\Text(130,5)[l]{$\nu$}
	\Text(25,35)[b]{$\bar{s}$}
	\Text(25,15)[t]{$u$}
	\Text(149,25)[r]{.}
    \end{picture}   \end{center}

The resolution to this fatal problem was put forward by Glashow, 
Iliopoulos, and Maiani.\cite{GIM}  Expand the model of quarks to 
include two left-handed doublets,
\begin{equation}
		\left(
		\begin{array}{c}
			\nu_{e}  \\
			e^{-}
		\end{array}
		 \right)_{L} \;\;
		\left(
		\begin{array}{c}
			\nu_{\mu}  \\
			\mu^{-}
		\end{array}
		 \right)_{L} 
  \;\;\;\;\;\;
 		\left(
		\begin{array}{c}
			u  \\
			d_{\theta}
		\end{array}
		 \right)_{L} \;\;
		\left(
		\begin{array}{c}
			c  \\
			s_{\theta}
		\end{array}
		 \right)_{L}  \; ,  \label{eq:gim}
\end{equation}
where 
\begin{equation}
    s_{\theta} = s\,\cos\theta_{C} - d\,\sin\theta_{C}\; ,
    \label{eq:sthdef}
\end{equation}
plus the corresponding right-handed singlets, $e_{R}$, $\mu_{R}$, 
$u_{R}$, $d_{R}$, $c_{R}$, and $s_{R}$.  This required the 
introduction of the charmed quark, $c$, which had not yet been 
observed.  By the addition of the second quark generation, the 
flavor-changing cross terms vanish in the $Z$-quark interaction, and 
we are left with:    \begin{center} \begin{picture}(280,100)(0,0)
	\ArrowLine(30,50)(10,90)
	\Text(5,85)[]{$q_{i}$}
	\ArrowLine(10,10)(30,50)
	\Text(5,15)[]{$q_{i}$}
	\ZigZag(30,50)(90,50){3}{6}
	\Text(95,58)[]{$\lambda$}
	\Text(110,50)[l]{${\displaystyle\frac{-ig}{4\cos\theta_{W}}}\gamma_{\lambda}[(1 - 
	\gamma_{5})L_{i} + (1 + \gamma_{5})R_{i}]$\quad ,}

    \end{picture}   \end{center}
which is flavor diagonal!

The generalization to $n$ quark doublets is straightforward.  Let the 
charged-current interaction be
\begin{equation}
    \lag_{W\mathrm{-quark}} = \frac{-g}{2\sqrt{2}}\left[\bar{\Psi}\gamma^{\mu}
    (1 - \gamma_{5}){\mathcal O}\Psi\:W^{+}_{\mu} + \mathrm{h.c.} \right]\; ,
    \label{eq:compWq}
\end{equation}
where the composite quark spinor is 
\begin{equation}
    \Psi = \left( 
    \begin{array}{c}
        u  \\
        c  \\
        \vdots  \\
           \\
        d  \\
        s  \\
        \vdots
    \end{array}
    \right)
    \label{eq:compspin}
\end{equation}
and the flavor structure is contained in
\begin{equation}
    {\mathcal O} = \left( 
    \begin{array}{cc}
        0 & U  \\
        0 & 0
    \end{array}
    \right)\; ,
    \label{eq:flav}
\end{equation}
where $U$ is the unitary quark-mixing matrix.  The weak-isospin 
contribution to the neutral-current interaction has the form
\begin{equation}
    \lag_{Z\mathrm{-quark}}^{\mathrm{iso}} = \frac{-g}{4\cos\theta_{W}}
    \bar{\Psi}\gamma^{\mu}(1 - \gamma_{5})\left[{\mathcal O},{\mathcal 
    O}^{\dagger}\right]\Psi\; .
    \label{eq:isoZq}
\end{equation}
Since the commutator
\begin{equation}
    \left[{\mathcal O},{\mathcal O}^{\dagger}\right] =
    \left(
    \begin{array}{cc}
        I & 0  \\
        0 & -I
    \end{array}
    \right)
    \label{eq:commut}
\end{equation}
the neutral-current interaction is flavor diagonal, and the 
weak-isospin piece is, as expected, proportional to $\tau_{3}$.

In general, the $n \times n$ quark-mixing matrix $U$ can be 
parametrized in terms of $n(n-1)/2$ real mixing angles and 
$(n-1)(n-2)/2$ complex phases, after exhausting the freedom to 
redefine the phases of quark fields.  The $3\times 3$ case, of three 
mixing angles and one phase, often called the 
Cabibbo--Kobayashi-Maskawa matrix, presaged the discovery of the third 
generation of quarks and leptons.\cite{KM}

\section{Precision Tests of the Electroweak Theory{\protect\cite{Y2K}}}
\subsection{Measurements on the $Z^{0}$ pole \label{subsec:Zpole}}
In its 
simplest form, with the electroweak gauge symmetry broken by the Higgs 
mechanism, the $SU(2)_{L}\otimes U(1)_{Y}$ theory has scored many 
qualitative successes: the prediction of neutral-current interactions, 
the necessity of charm, the prediction of the existence and properties 
of the weak bosons $W^{\pm}$ and $Z^{0}$.  Over the past ten years, in 
great measure due to the beautiful experiments carried out at the $Z$ 
factories at CERN and SLAC, precision measurements have tested the
electroweak theory as a quantum field theory,\cite{sirlin,morris} at
the one-per-mille level, as indicated in Table \ref{tbl:Zmeas}.  A
classic achievement of the $Z$ factories is the determination of the
number of light neutrino species.  If we define the invisible width
\begin{table}[b!]
    \centering
    \caption{Precision measurements at the $Z^{0}$ pole. (For 
    sources of the data, see the Review of Particle Physics, 
    Ref.\ {\protect\cite{pdg}} and the LEP Electroweak Working Group 
    web server, {\protect \url{http://www.cern.ch/LEPEWWG/}}.)}
    \begin{tabular}{cc}
        \hline\\[-6pt]
        $M_{Z}$ & $91\,188 \pm 2.2\mevcc$ \\
        %\hline
        $\Gamma_{Z}$ & $2495.2 \pm 2.6\mev$  \\
        %\hline
        $\sigma^{0}_{\mathrm{hadronic}}$ & $41.541 \pm 0.037\nb$  \\
        %\hline
        $\Gamma_{\mathrm{hadronic}}$ & $1743.8 \pm 2.2\mev$  \\
        %\hline
        $\Gamma_{\mathrm{leptonic}}$ & $84.057 \pm 0.088\mev$  \\
        %\hline
        $\Gamma_{\mathrm{invisible}}$ & $499.4 \pm 1.7\mev$  \\[4pt]
        \hline
    \end{tabular}
    \label{tbl:Zmeas}
\end{table}
of the $Z^{0}$ as 
\begin{equation}
    \Gamma_{\mathrm{invisible}} = \Gamma_{Z} - 
    \Gamma_{\mathrm{hadronic}} - 
    3\Gamma_{\mathrm{leptonic}}\; ,
    \label{eq:gaminv}
\end{equation}
then we can compute the number of light neutrino species as
\begin{equation}
    N_{\nu} = \Gamma_{\mathrm{invisible}}/\Gamma^{\mathrm{SM}}(Z \to 
    \nu_{i}\bar{\nu}_{i})\;.
    \label{eq:nnudef}
\end{equation}
A typical current value is $N_{\nu} = 2.984 \pm 0.008$, in excellent 
agreement with the observation of light $\nu_{e}$, $\nu_{\mu}$, and 
$\nu_{\tau}$.

\begin{figure}[tb]
	\centerline{\BoxedEPSF{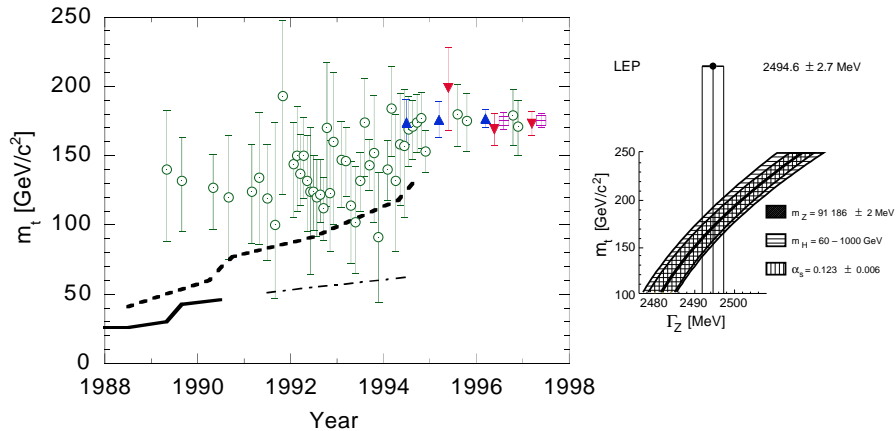  scaled 450}}
\caption{Indirect determinations of the top-quark mass from fits to 
electroweak observables (open circles) and 95\% confidence-level lower 
bounds on the top-quark mass inferred from direct searches in 
$e^{+}e^{-}$ annihilations (solid line) and in $\bar{p}p$ collisions, 
assuming that standard decay modes dominate (broken line).  An 
indirect lower bound, derived from the $W$-boson width inferred from 
$\bar{p}p \rightarrow (W\hbox{ or }Z)+\hbox{ anything}$, is shown as 
the dot-dashed line.  Direct measurements of $m_{t}$ by the CDF 
(triangles) and D\O\ (inverted triangles) Collaborations are shown at 
the time of initial evidence, discovery claim, and at the conclusion 
of Run 1.  The world average from direct observations is shown as the
crossed box.  For sources of data, see Ref.  {\protect\cite{pdg}}. 
\textit{Inset:} Electroweak theory predictions for the width of the
$Z^{0}$ boson as a function of the top-quark mass, compared with the
width measured in LEP experiments.  (From Ref.\
{\protect\cite{cqpt}}.)}
	\label{EWtop}
\end{figure}

As an example of the insights precision 
measurements have brought us (one that mightily impressed the Royal 
Swedish Academy of Sciences in 1999), I show in Figure \ref{EWtop} the time 
evolution of the top-quark mass favored by simultaneous fits to many 
electroweak observables.  Higher-order processes involving virtual top 
quarks are an important element in quantum corrections to the 
predictions the electroweak theory makes for many observables.  A case 
in point is the total decay rate, or width, of the $Z^{0}$ boson: the 
comparison of experiment and theory shown in the inset to Figure 
\ref{EWtop} favors a top mass in the neighborhood of $180\gevcc$.

\begin{figure}[tb] 
\centerline{\BoxedEPSF{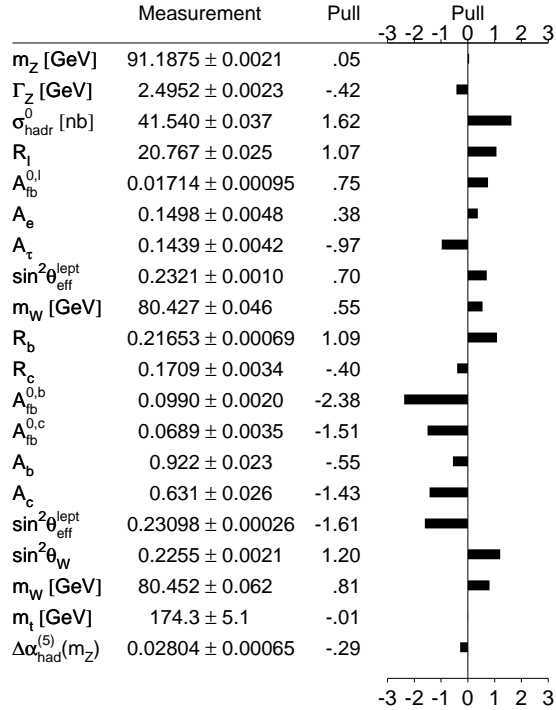 scaled 375}}
\vspace{10pt}
\caption{Precision electroweak measurements and the pulls they exert 
on a global fit to the standard model, from Ref.\ 
{\protect\cite{ewwg}}.}
\label{fig:pulls}
\end{figure}

The comparison between the electroweak theory and a considerable 
universe of data is shown in Figure \ref{fig:pulls},\cite{ewwg} 
where the pull, or difference between the global fit and measured 
value in units of standard deviations, is shown for some twenty 
observables.
The distribution of pulls for this fit, due to the LEP Electroweak 
Working Group, is not noticeably different from a normal 
distribution, and only a couple of observables differ from the fit by 
as much as about two standard deviations.  This is the case for any 
of the recent fits.  From fits of the kind represented here, 
we learn that the standard-model interpretation of the data favors a 
light Higgs boson.\cite{wjmsf}  We will revisit this conclusion in 
\S\ref{Hclues}.

The beautiful agreement between the electroweak theory and a vast 
array of data from neutrino interactions, hadron collisions, and 
electron-positron annihilations at the $Z^{0}$ pole and beyond means 
that electroweak studies have become a modern arena in which we can 
look for new physics ``in the sixth place of 
decimals.''

\subsection{Parity violation in atomic physics}
Later in TASI00, you will hear from Carl Wieman\cite{CWtasi} about the beautiful 
table-top experiments that probe the structure of the weak neutral 
current.  I want to spend a few minutes laying the foundation for the 
interpretation of those measurements.\footnote{The book by Commins 
and Bucksbaum, Ref.\ \cite{combuck}, is an excellent reference for 
many ``real-world'' problems at low $Q^{2}$.}

At very low momentum transfers, as in atomic physics applications, 
the nucleon appears elementary, and so we can write an effective 
Lagrangian for nucleon $\beta$ decay in the limit of zero momentum 
transfer as
\begin{equation}
    \lag_{\beta} = -\;\frac{G_{F}}{\sqrt{2}} \bar{e}\gamma_{\lambda} 
    (1 - \gamma_{5}) \nu\, \bar{p} \gamma^{\lambda} (1 - 
    g_{A}\gamma_{5}) n \; ,
    \label{eq:betadklag}
\end{equation}
where $g_{A} \approx 1.26$ is the axial charge of the nucleon.  
Accordingly, the neutral-current interactions are
\begin{eqnarray}
    \lag{ep} & = & \frac{G_{F}}{2\sqrt{2}} \bar{e} \gamma_{\lambda}
    (1 - 4x_{W} - \gamma_{5}) e\; \bar{p} \gamma^{\lambda}
    (1 - 4x_{W} - \gamma_{5}) p \; , \nonumber \\
    \lag{en} & = & \frac{G_{F}}{2\sqrt{2}} \bar{e} \gamma_{\lambda}
    (1 - 4x_{W} - \gamma_{5}) e\; \bar{n} \gamma^{\lambda}
    (1 - \gamma_{5}) n \; ,
    \label{eq:AeN}
\end{eqnarray}
where $x_{W} \equiv \sin^{2}\theta_{W}$.

For our purposes, we may regard the nucleus as a noninteracting 
collection of $Z$ protons and $N$ neutrons.  We perform a 
nonrelativistic reduction of the interaction; nucleons contribute 
coherently to the axial-electron--vector-nucleon ($A_{e}V_{N}$) 
coupling, so the dominant parity-violating contribution to the $eN$ 
amplitude is
\begin{equation}
    \M_{\mathrm{p.v.}} = \frac{-iG_{F}}{2\sqrt{2}} Q^{W} \bar{e}
    \rho_{N}(\mathbf{r})\gamma_{5} e \; ,
    \label{eq:APpv}
\end{equation}
where $\rho_{N}(\mathbf{r})$ is the nucleon density at the electron 
coordinate $\mathbf{r}$, and $Q^{W} \equiv Z(1-4x_{W})-N$ is the weak 
charge.

Bennett and Wieman (Boulder) have reported a new determination of the 
weak charge of Cesium by measuring the transition polarizability for the 
6S-7S transition.\cite{benwie}  The new value,
\begin{equation}
	Q_W(\textrm{Cs})= -72.06 \pm 0.28\hbox{ (expt)} \pm 0.34\hbox{ 
	(theory),} 
	\label{eq:wieman}
\end{equation}
represents a sevenfold improvement in the experimental error and a 
significant reduction in the theoretical uncertainty.  It lies about
2.5 standard deviations above the prediction of the standard model.  We 
are left with the traditional situation in which elegant measurements 
of parity nonconservation in atoms are on the edge of incompatibility 
with the standard model.

A number of authors\cite{erlang} have noted that the discrepancy in
the weak charge $Q_{W}$ and a 2-$\sigma$ anomaly\footnote{In Erler \&
Langacker's fit, for example, the number of light neutrino species
inferred from the invisible width of the $Z^{0}$ is $N_{\nu} = 2.985
\pm 0.008 = 3 - 2\sigma$.} in the total width of the $Z^{0}$ can be
reduced by introducing a $Z^{\prime}$ boson with a mass of about
$800\gevcc$.  The additional neutral gauge boson resembles the
$Z_{\chi}$ familiar from unified theories based on the group 
$E_{6}$.\cite{Hewett:1989xc}  

\subsection{The vacuum energy problem} 
I want to spend a moment to revisit a 
longstanding, but usually unspoken, challenge to the completeness of 
the electroweak theory as we have defined it: the vacuum energy 
problem.\cite{Veltman:1975au}
I do so not only for its intrinsic interest, but also to 
raise the question, ``Which problems of completeness and 
consistency do we worry about at a given moment?''  It is perfectly 
acceptable science---indeed, it is often essential---to put certain 
problems aside, in the expectation that we will return to them at the 
right moment.  What is important is never to forget that the problems 
are there, even if we do not allow them to paralyze us.  

For the usual Higgs potential, 
$V(\varphi^{\dagger}\varphi) = \mu^{2}(\varphi^{\dagger}\varphi) + 
\abs{\lambda}(\varphi^{\dagger}\varphi)^{2}$, the value of 
the potential at the minimum is
\begin{equation}
    V(\vev{\varphi^{\dagger}\varphi}) = \frac{\mu^{2}v^{2}}{4} = 
    - \frac{\abs{\lambda}v^{4}}{4} < 0.
    \label{minpot}
\end{equation}
Identifying $M_{H}^{2} = -2\mu^{2}$, we see that the Higgs potential 
contributes a field-independent constant term,
\begin{equation}
    \varrho_{H} \equiv \frac{M_{H}^{2}v^{2}}{8}.
    \label{eq:rhoH}
\end{equation}
I have chosen the notation $\varrho_{H}$ because the constant term in the 
Lagrangian plays the role of a vacuum energy density.  When we 
consider gravitation, adding a vacuum energy density 
$\varrho_{\mathrm{vac}}$ is equivalent to adding a cosmological constant 
term to Einstein's equation.  Although recent
observations\footnote{For a cogent summary of current knowledge of the
cosmological parameters, including evidence for a cosmological
constant, see Ref.\ \cite{cosconst}.} raise the intriguing possibility
that the cosmological constant may be different from zero, the
essential observational fact is that the vacuum energy density must be
very tiny indeed,\footnote{For a useful summary of gravitational
theory, see the essay by T. d'Amour in \S14 of the 2000 \textit{Review
of Particle Physics,} Ref.\ \cite{pdg}.}
\begin{equation}
    \varrho_{\mathrm{vac}} \ltap 10^{-46}\gev^{4}\; .
    \label{eq:rhovaclim}
\end{equation}
Therein lies the puzzle: if we take
$v = (G_F\sqrt{2})^{-\frac{1}{2}}  \approx 246\gev$  
and insert the current experimental lower bound\cite{higgslim2k} 
$M_{H} \gtap 113.5\gevcc$ into \eqn{eq:rhoH}, we find that the 
contribution of the Higgs field to the vacuum energy density is
\begin{equation}
    \varrho_{H} \gtap  10^{8}\gev^{4},
    \label{eq:rhoHval}
\end{equation}
some 54 orders of magnitude larger than the upper bound inferred from 
the cosmological constant.

What are we to make of this mismatch?  The fact that $\varrho_{H} \gg 
\varrho_{\mathrm{vac}}$ means that the smallness of the cosmological 
constant needs to be explained.  In a unified theory of the strong, 
weak, and electromagnetic interactions, other (heavy!) Higgs fields 
have nonzero vacuum expectation values that may give rise to still 
greater mismatches.  At a fundamental level, we can therefore conclude 
that a spontaneously broken gauge theory of the strong, weak, and 
electromagnetic interactions---or merely of the electroweak 
interactions---cannot be complete.  Either we must find a separate 
principle to zero the vacuum energy density of the Higgs field, or 
we may suppose that a proper quantum theory of gravity, in combination 
with the other interactions, will resolve the puzzle of the 
cosmological constant.  The vacuum energy problem must be an important 
clue.  But to what?

%\clearpage

\section{The Higgs Boson}
\subsection{Why the Higgs boson must exist}
How can we be sure that a Higgs boson, or something very like it, will be 
found? One 
path to the \emph{theoretical} discovery of the Higgs boson
involves its role in the cancellation of 
high-energy divergences. An illuminating example is provided by the 
reaction
\begin{figure}[tb]
	\centerline{\BoxedEPSF{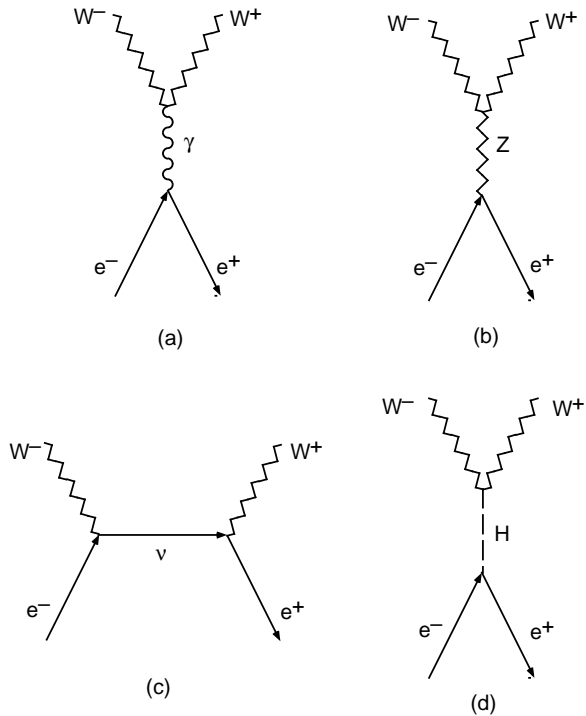  scaled 700}}
	\vspace*{6pt}
	\caption{Lowest-order contributions to the $e^+e^- \rightarrow 
	W^{+}W^{-}$ scattering amplitude.}
	\protect\label{fig:eeWW}
\end{figure}
\begin{equation}
	e^+e^- \to W^+W^- \; ,
\end{equation}
which is described in lowest order by the four 
Feynman graphs in Figure \ref{fig:eeWW}. The contributions of the direct-channel 
$\gamma$- and $Z^0$-exchange diagrams 
of Figs.~\ref{fig:eeWW}(a) and (b) cancel the leading divergence in the $J=1$ 
partial-wave amplitude of 
the neutrino-exchange diagram in Figure~\ref{fig:eeWW}(c).  This is 
the famous ``gauge cancellation'' observed in experiments at LEP~2 
and the Tevatron.  The LEP measurements in Figure~\ref{fig:LEPgc}
\begin{figure}[tb]
	\centerline{\BoxedEPSF{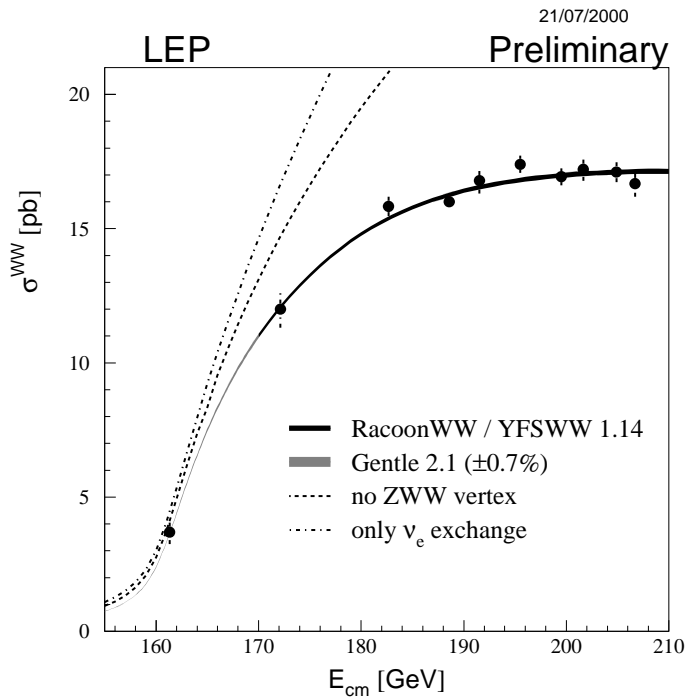  scaled 500}}
	\vspace*{6pt}
	\caption{Cross section for the reaction $e^{+}e^{-} \to W^{+}W^{-}$ 
	measured by the four LEP experiments, together with the full 
	electroweak-theory simulation and the cross sections that would 
	result from $\nu$-exchange alone and from $(\nu+\gamma)$-exchange
	(from {\protect \url{http://www.cern.ch/LEPEWWG/}}).}
	\protect\label{fig:LEPgc}
\end{figure}
agree well with the predictions of electroweak-theory Monte Carlo
generators, which predict a benign high-energy behavior.  If the 
$Z$-exchange contribution is omitted (dashed line) or if both the 
$\gamma$- and $Z$-exchange contributions are omitted (dot-dashed 
line), the calculated cross section grows unacceptably with 
energy---and disagrees with the measurements.  The gauge cancellation 
in the $J=1$ partial-wave amplitude is thus observed.

However, this is not the end of the high-energy story: the $J=0$
partial-wave amplitude, which exists in this case because the
electrons are massive and may therefore be found in the ``wrong''
helicity state, grows as $s^{1/2}$ for the production of
longitudinally polarized gauge bosons.  The resulting divergence is
precisely cancelled by the Higgs boson graph of
Figure~\ref{fig:eeWW}(d).  If the Higgs boson did not exist, something
else would have to play this role.  From the point of view of
$S$-matrix analysis, the Higgs-electron-electron coupling must be
proportional to the electron mass, because ``wrong-helicity''
amplitudes are always proportional to the fermion mass.

Let us underline this result.
If the gauge symmetry were unbroken, there would be 
no Higgs boson, no longitudinal gauge bosons, and no extreme divergence 
difficulties. But there would be no viable low-energy phenomenology
 of the 
weak interactions. The most severe divergences of individual diagrams 
are eliminated by the gauge 
structure of the couplings among gauge bosons and leptons. A lesser, but 
still potentially fatal, divergence arises because the electron has 
acquired mass---because of the Higgs mechanism. Spontaneous symmetry 
breaking provides its own cure by supplying a Higgs boson to remove the 
last divergence. A similar interplay and compensation must exist in any 
satisfactory theory.

\subsection{Bounds on $M_{H}$}
The Standard Model does not give a precise 
prediction for the mass of the Higgs boson. We can, however, use arguments 
of self-consistency to place plausible lower and upper bounds on the mass of 
the Higgs particle in the minimal model. Unitarity arguments\cite{lqt} lead to a conditional upper bound on the Higgs 
boson mass. It is straightforward to compute the 
amplitudes ${\cal M}$ for gauge boson scattering at high energies, and to make
a partial-wave decomposition, according to
\begin{equation}
      {\cal M}(s,t)=16\pi\sum_J(2J+1)a_J(s)P_J(\cos{\theta}) \; .
\end{equation}
 Most channels ``decouple,'' in the sense 
that partial-wave amplitudes are small at all energies (except very
near the particle poles, or at exponentially large energies), for
any value of the Higgs boson mass $M_H$. Four channels are interesting:
\begin{equation}
\begin{array}{cccc}
W_L^+W_L^- & Z_L^0Z_L^0/\sqrt{2} & HH/\sqrt{2} & HZ_L^0 \; ,
\end{array}
\end{equation}
where the subscript $L$ denotes the longitudinal polarization
states, and the factors of $\sqrt{2}$ account for identical particle
statistics. For these, the $s$-wave amplitudes are all asymptotically
constant (\ie, well-behaved) and  
proportional to $G_FM_H^2.$ In the high-energy 
limit,\footnote{It is convenient to calculate these amplitudes by 
means of the Goldstone-boson equivalence theorem,\cite{EQT} which 
reduces the dynamics of longitudinally polarized gauge bosons to a 
scalar field theory with interaction Lagrangian given by 
$\mathcal{L}_{\mathrm{int}} = -\lambda v h 
(2w^{+}w^{-}+z^{2}+h^{2}) - 
(\lambda/4)(2w^{+}w^{-}+z^{2}+h^{2})^{2}$, with $1/v^{2} = 
G_{F}\sqrt{2}$ and $\lambda = G_{F}M_{H}^{2}/\sqrt{2}$.}
\begin{equation}
\lim_{s\gg M_H^2}(a_0)\to\frac{-G_F M_H^2}{4\pi\sqrt{2}}\cdot \left[
\begin{array}{cccc} 1 & 1/\sqrt{8} & 1/\sqrt{8} & 0 \\
      1/\sqrt{8} & 3/4 & 1/4 & 0 \\
      1/\sqrt{8} & 1/4 & 3/4 & 0 \\
      0 & 0 & 0 & 1/2 \end{array} \right] \; .
\end{equation} 
Requiring that the largest eigenvalue respect the 
partial-wave unitarity condition $\abs{a_0}\le 1$ yields
\begin{equation}
	M_H \le \left(\frac{8\pi\sqrt{2}}{3G_F}\right)^{1/2} =1\tevcc
\end{equation}
as a condition for perturbative unitarity.

If the bound is respected, weak interactions remain weak at all
energies, and perturbation theory is everywhere reliable. If the
bound is violated, perturbation theory breaks down, and weak
interactions among $W^\pm$, $Z$, and $H$ become strong on the \onetev.
This means that the features of strong interactions at GeV energies
will come to characterize electroweak gauge boson interactions at
TeV energies. We interpret this to mean that new phenomena are to
be found in the electroweak interactions at energies not much larger
than 1~TeV.

It is worthwhile to note in passing that 
the threshold behavior of the partial-wave amplitudes for gauge-boson 
scattering follows generally from chiral symmetry.\cite{LT8}  The partial-wave 
amplitudes $a_{IJ}$ of definite isospin $I$ and angular momentum $J$ are 
given by
\begin{eqnarray}
	a_{00} \approx & G_Fs/8\pi\sqrt{2} & \hbox{attractive,}
	\nonumber \\
	a_{11}  \approx & G_Fs/48\pi\sqrt{2} & \hbox{attractive,} \\
	a_{20} \approx & -G_Fs/16\pi\sqrt{2} & \hbox{repulsive.}
	\nonumber
\end{eqnarray} 

The electroweak theory itself provides another reason to expect that 
discoveries will not end with the Higgs boson.  Scalar field theories 
make sense on all energy scales only if they are noninteracting, or 
``trivial.''\cite{15}  The vacuum of quantum field theory is a dielectric 
medium that screens charge.  Accordingly, the effective charge is a 
function of the distance or, equivalently, of the energy scale.  This is 
the famous phenomenon of the running coupling constant.

In $\lambda\phi^4$ theory (compare the interaction term in the Higgs 
potential), it is easy to calculate the variation of the coupling 
constant $\lambda$ in perturbation theory by summing bubble graphs like 
this one:
\begin{equation}
\BoxedEPSF{Bullex.epsf  scaled 600}\;\;\;\;.
\end{equation} \vphantom{{\LARGE |}}The coupling constant $\lambda(\mu)$ on a physical scale $\mu$ 
is related 
to the coupling constant on a higher scale $\Lambda$ by
\begin{equation}
\frac{1}{\lambda(\mu)} = \frac{1}{\lambda(\Lambda)} + 
\frac{3}{2\pi^2}\log{\left(\Lambda/\mu\right)}\;\;.
\label{rng}
\end{equation}
This perturbation-theory result is reliable only when $\lambda$ is small, 
but lattice field theory allows us to treat the strong-coupling regime.

In order for the Higgs potential to be stable (\ie, for the energy of the 
vacuum state not to race off to $-\infty$), $\lambda(\Lambda)$ must not 
be negative.  Therefore we can rewrite \eqn{rng} as an inequality,
\begin{equation}
\frac{1}{\lambda(\mu)} \ge 
\frac{3}{2\pi^2}\log{\left(\Lambda/\mu\right)}\;\;. 
\end{equation}
This gives us an {\em upper bound},
\begin{equation}
\lambda(\mu) \le 
2\pi^2/3\log{\left(\Lambda/\mu\right)}\;\;,
\label{upb}
\end{equation}
on the coupling strength at the physical scale $\mu$.
If we require the theory to make sense to arbitrarily high energies---or 
short distances---then we must take the limit $\Lambda\rightarrow\infty$ 
while holding $\mu$ fixed at some reasonable physical scale.  In this 
limit, the bound \eqn{upb} forces $\lambda(\mu)$ to zero.  The scalar field 
theory has become free field theory; in theorist's jargon, it is trivial.

We can rewrite the inequality \eqn{upb} as a bound on the Higgs-boson mass.  
Rearranging and exponentiating both sides gives the condition
\begin{equation}
\Lambda \le \mu \exp{\left(\frac{2\pi^2}{3\lambda(\mu)}\right)}\;\;.
\end{equation}
Choosing the physical scale as $\mu=M_H$, and remembering that, before 
quantum corrections,
\begin{equation}
M_H^2 = 2\lambda(M_H)v^2\;\;,
\end{equation}
where $v=(G_F\sqrt{2})^{-1/2}\approx 246~\hbox{GeV}$ is the vacuum 
expectation value of the Higgs field times $\sqrt{2}$, we find that
\begin{equation}
\Lambda \le M_H\exp{\left(\frac{4\pi^2v^2}{3M_H^2}\right)}\;\;.
\end{equation}
For any given Higgs-boson mass, there is a maximum energy scale 
$\Lambda^\star$ at which the theory ceases to make sense.  The 
description of the Higgs boson as an elementary scalar is at best an 
effective theory, valid over a finite range of energies.

\begin{figure}[tb]
	\centerline{\BoxedEPSF{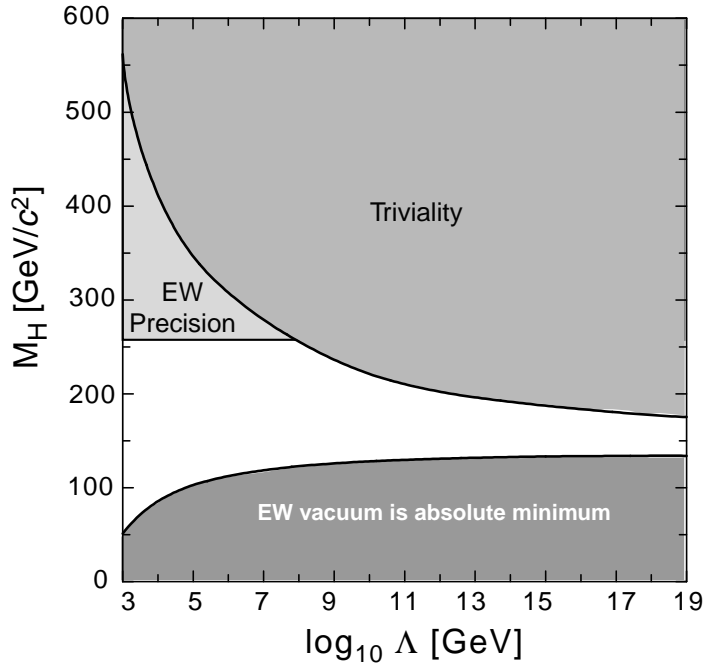  scaled 700}}
	\vspace*{6pt}
	\caption{Bounds on the Higgs-boson mass that follow from 
	requirements that the electroweak theory be consistent up to the 
	energy scale $\Lambda$.  The upper bound follows from triviality 
	conditions; the lower bound follows from the requirement that $V(v) < 
	V(0)$.  Also shown is the range of masses permitted at the 95\%\ 
	confidence level by precision measurements.}
	\protect\label{fig:Hbds}
\end{figure}

This perturbative analysis breaks down when the Higgs-boson mass 
approaches $1~\hbox{TeV}\!/\!c^2$ and the interactions become strong.  
Lattice analyses\cite{17} indicate that, for the theory to describe 
physics to an accuracy of a few percent up 
to a few TeV, the mass of the Higgs boson can be no more than about 
$710\pm 60\gevcc$.  Another way of putting this result is that, if 
the elementary Higgs boson takes on the largest mass allowed by 
perturbative unitarity arguments, the electroweak theory will be living 
on the brink of instability.

A lower bound is obtained by 
computing\cite{SSI18} the first quantum corrections to the classical potential
\eqn{SSBpot}. Requiring that $\vev{\phi}\neq 0$ be an absolute minimum of the one-loop 
potential up to a scale $\Lambda$ yields the vacuum-stability condition 
\begin{equation}
	M_H^2 > \frac{3G_F\sqrt{2}}{8\pi^{2}}(2M_W^4+M_Z^4-4m_{t}^{4})
	\log(\Lambda^{2}/v^{2}) \; .
\end{equation}

The upper and lower bounds plotted in Figure \ref{fig:Hbds} are the results of 
full two-loop calculations.\cite{2loopvacstab}  There I have also 
indicated the upper bound on $M_{H}$ derived from precision 
electroweak measurements \textit{in the framework of the standard electroweak 
theory.}  If the Higgs boson is relatively light---which would itself require 
explana\-tion---then the theory can be self-consistent up to 
very high energies.  If the electroweak theory is to make sense all the 
way up to a unification scale $\Lambda^\star = 10^{16}~\hbox{GeV}$, then 
the Higgs-boson mass must lie in the interval $134\gevcc \ltap M_{W}
\ltap 177 \gevcc$.\cite{16}

\subsection{Higgs-Boson Properties}
Once we assume a value for the Higgs-boson mass, it is a simple matter 
to compute the rates for Higgs-boson decay into pairs of fermions or 
weak bosons.\cite{Ellis:1976ap}  For a fermion with color $N_{c}$, the partial width is
\begin{equation}
	\Gamma(H \to f\bar{f}) = \frac{G_{F}m_{f}^{2}M_{H}}{4\pi\sqrt{2}} 
	\cdot N_{c} \cdot \left( 1 - \frac{4m_{f}^{2}}{M_{H}^{2}} 
	\right)^{3/2} \; ,
	\label{eq:Higgsff}
\end{equation}
which is proportional to $M_{H}$ in the limit of large Higgs mass.
The partial width for decay into a $W^{+}W^{-}$ pair is
\begin{equation}
	\Gamma(H \to W^{+}W^{-}) = \frac{G_{F}M_{H}^{3}}{32\pi\sqrt{2}} 
	(1 - x)^{1/2} (4 -4x +3x^{2}) \; ,
	\label{eq:HiggsWW}
\end{equation}
where $x \equiv 4M_{W}^{2}/M_{H}^{2}$.  Similarly, the partial width 
for decay into a pair of $Z^{0}$ bosons is 
\begin{equation}
	\Gamma(H \to Z^{0}Z^{0}) = \frac{G_{F}M_{H}^{3}}{64\pi\sqrt{2}} 
	(1 - x^{\prime})^{1/2} (4 -4x^{\prime} +3x^{\prime 2}) \; ,
	\label{eq:HiggsZZ}
\end{equation}
where $x^{\prime} \equiv 4M_{Z}^{2}/M_{H}^{2}$.  The rates for decays into 
weak-boson pairs are asymptotically proportional to $M_{H}^{3}$ and 
$\cfrac{1}{2}M_{H}^{3}$, respectively, the factor $\cfrac{1}{2}$ 
arising from weak isospin.  In the final factors of \eqn{eq:HiggsWW} 
and \eqn{eq:HiggsZZ}, $2x^{2}$ and $2x^{\prime 2}$, respectively, 
arise from decays into transversely polarized gauge bosons.  The 
dominant decays for large $M_{H}$ are into pairs of longitudinally 
polarized weak bosons.

\begin{figure}[tbh]
	\centerline{\BoxedEPSF{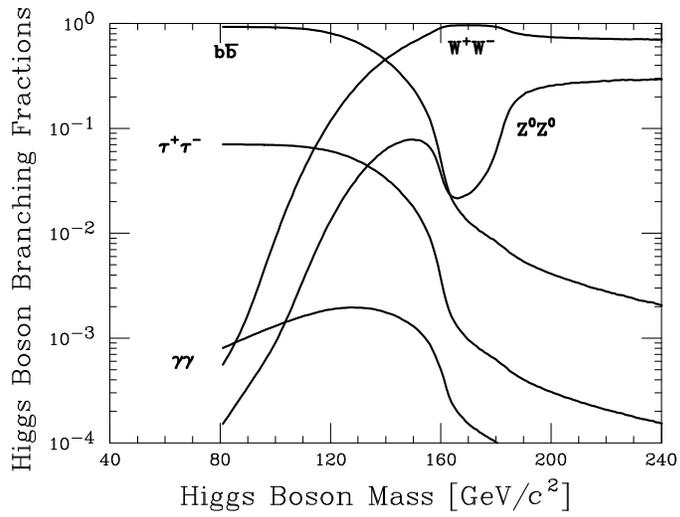  scaled 800}}
	\vspace*{6pt}
	\caption{Branching fractions for the prominent decay modes of a light 
	Higgs boson.}
	\protect\label{fig:LHdk}
\end{figure}

Branching fractions for decay modes that hold promise for the 
detection of a light Higgs boson are displayed in Figure 
\ref{fig:LHdk}.  In addition to the $f\bar{f}$ and $VV$ modes that 
arise at tree level, I have included the $\gamma\gamma$ mode that 
proceeds through loop diagrams.  Though rare, the $\gamma\gamma$ 
channel offers an important target for LHC experiments.

\begin{figure}[t!]
	\centerline{\BoxedEPSF{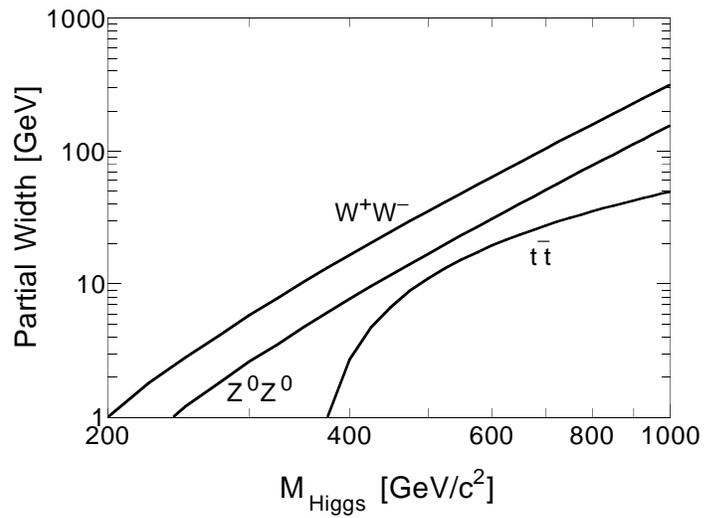  scaled 800}}
	\vspace*{6pt}
	\caption{Partial widths for the prominent decay modes of a heavy Higgs 
	boson.}
	\protect\label{fig:HHdk}
\end{figure}

Figure \ref{fig:HHdk} shows the partial widths for the decay of a 
Higgs boson into the dominant $W^+W^-$ and $Z^0Z^0$ channels and into 
$t\bar{t}$, for $m_t = 175\gevcc$.  Whether the $t\bar{t}$ mode will 
be useful to confirm the observation of a heavy Higgs boson, or merely 
drains probability from the $ZZ$ channel favored for a heavy-Higgs 
search, is a question for detailed detector simulations.

\begin{figure}[t!]
	\centerline{\BoxedEPSF{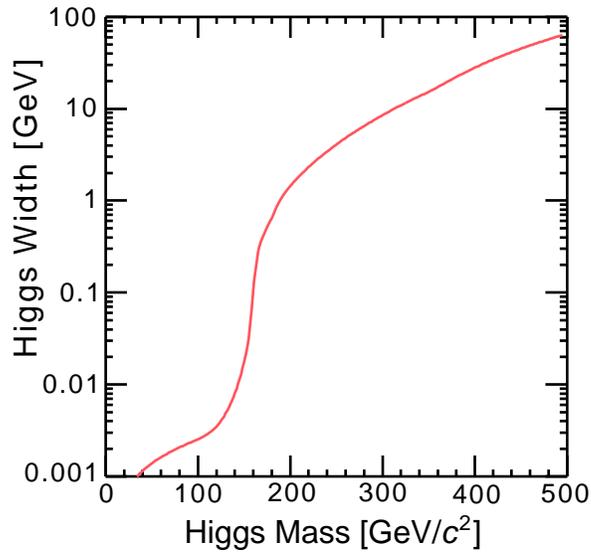  scaled 1000}}
	\vspace*{6pt}
	\caption{Higgs-boson total width as a function of mass.}
	\protect\label{fig:Htot}
\end{figure}

Below the $W^{+}W^{-}$ threshold, the total width of the 
standard-model Higgs boson is rather small, typically less than 
$1\gev$.  Far above the threshold for decay into gauge-boson pairs, 
the total width is proportional to $M_{H}^{3}$.  At masses 
approaching $1\tevcc$, the Higgs boson is an ephemeron, with a 
perturbative width approaching its mass.  The Higgs-boson total width 
is plotted as a function of $M_{H}$ in Figure \ref{fig:Htot}.

\subsection{Clues to the Higgs-boson mass \label{Hclues}}
We have seen in \S\ref{subsec:Zpole} that the sensitivity of 
electroweak observables to the (long unknown) mass of the top quark 
gave early indications for a very massive top.  For example, the 
quantum corrections to the standard-model predictions \eqn{gbmsw} for 
$M_{W}$ and \eqn{gbmsz} for $M_{Z}$ arise from different quark loops:
\begin{center} \begin{picture}(280,80)(0,0)
	 \ZigZag(10,40)(50,40){2}{5}
	 \ZigZag(75,40)(115,40){2}{5}
	 \ArrowArc(62.5,40)(12.5,0,180)
	 \ArrowArc(62.5,40)(12.5,180,360)
	 \Text(62.5,58)[b]{{\large $\bar{b}$}}
	 \Text(62.5,22)[t]{{\large $t$}}
	 \Text(5,40)[r]{{\large $W^{+}$}}
	 \Text(120,40)[l]{{\large $W^{+}$}}
	 
	 \ZigZag(165,40)(205,40){2}{5}
	 \ZigZag(230,40)(270,40){2}{5}
	 \ArrowArc(217.5,40)(12.5,0,180)
	 \ArrowArc(217.5,40)(12.5,180,360)
	 \Text(217.5,58)[b]{{\large $\bar{t}$}}
	 \Text(217.5,22)[t]{{\large $t$}}
	 \Text(160,40)[r]{{\large $Z^{0}$}}
	 \Text(275,40)[l]{{\large $Z^{0}$,}}
    \end{picture}   \end{center}
$t\bar{b}$ for $M_{W}$, and $t\bar{t}$ (or $b\bar{b}$) for $M_{Z}$.  
These quantum corrections alter the link \eqn{gbmsz} between the $W$- and 
$Z$-boson masses, so that
\begin{equation}
    M_{W}^{2} = M_{Z}^{2}\left(1 - \sin^{2}\theta_{W}\right)
    \left(1 + \Delta\rho\right)\; ,
    \label{eq:MWqc}
\end{equation}
where
\begin{equation}
    \Delta\rho \approx \Delta\rho^{(\mathrm{quarks})} = 
    \frac{3G_{F}m_{t}^{2}}{8\pi^{2}\sqrt{2}} \; .
    \label{eq:drhoq}
\end{equation}
The strong dependence on $m_{t}^{2}$ is characteristic, and it 
accounts for the precision of the top-quark mass estimates derived 
from electroweak observables.

\begin{figure}[tb]
	\centerline{\BoxedEPSF{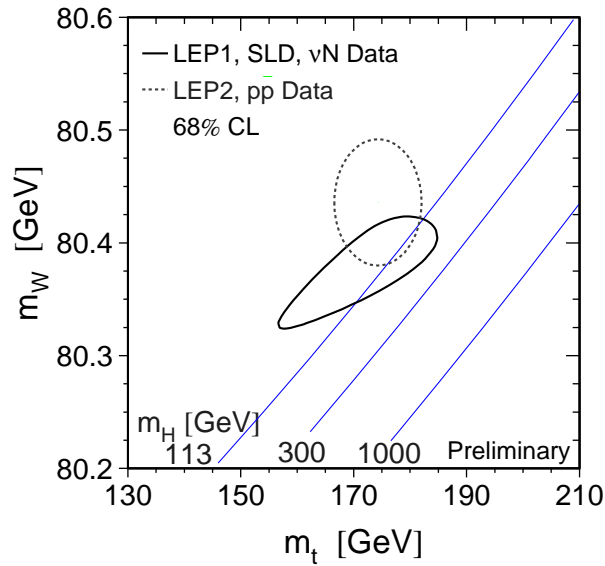  scaled 400}}
	\vspace*{6pt}
	\caption{Comparison of the indirect measurements of $M_{W}$ and
	$m_{t}$ (LEP~I+SLD+$\nu N$ data, solid contour) and the direct
	measurements (Tevatron and LEP~II data, dashed contour).  Also shown
	is the standard-model relationship for the masses as a function of the
	Higgs mass.  (From the LEP Electroweak Working Group, Ref.\
	{\protect\cite{ewwg}}.)} \protect\label{fig:LEPmwmt}
\end{figure}

Now that $m_{t}$ is known to about 3\% from direct observations at the
Tevatron, it becomes profitable to look beyond the quark loops to the
next most important quantum corrections, which arise from Higgs-boson
effects.  The Higgs-boson quantum corrections are typically smaller
than the top-quark corrections, and exhibit a more subtle dependence
on $M_{H}$ than the $m_{t}^{2}$ dependence of the top-quark
corrections.  For the case at hand,
\begin{equation}
    \Delta\rho^{\mathrm{(Higgs)}} = \mathcal{C} \cdot \ln \left(\frac{M_{H}}{v}\right) 
    \; ,
    \label{eq:drhoh}
\end{equation}
where I have arbitrarily chosen to define the coefficient
$\mathcal{C}$ at the electroweak scale $v$.  Since $M_{Z}$ has been
determined at LEP to 23~ppm, it is interesting to examine (see Figure
\ref{fig:LEPmwmt}) the dependence of $M_{W}$ upon $m_{t}$ and $M_{H}$.
\begin{figure}[tb]
	\centerline{\BoxedEPSF{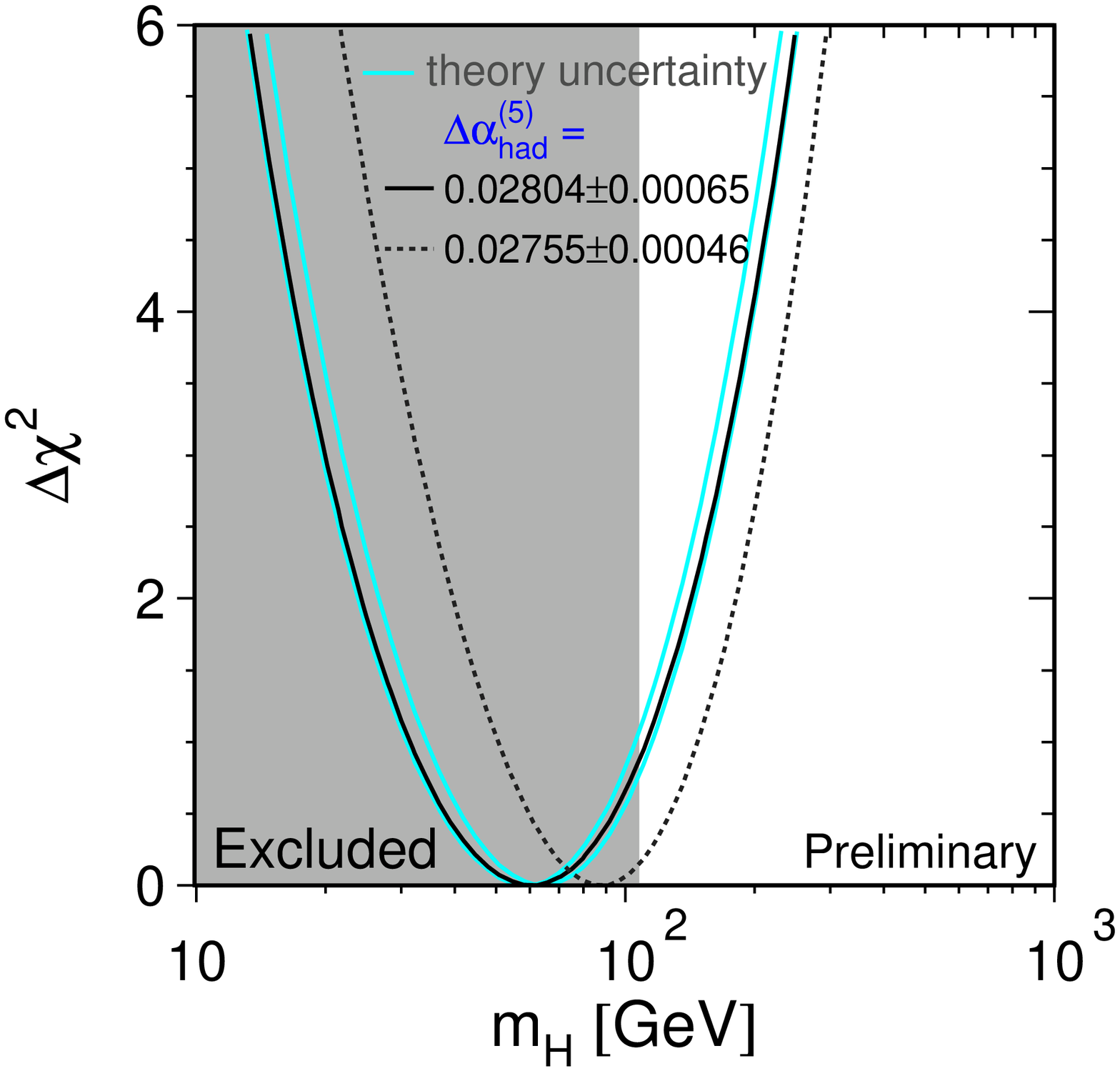  scaled 400}}
	\vspace*{6pt}
	\caption{$\Delta\chi^{2}=\chi^2-\chi^2_{\mathrm{min}}$ from a global fit to
	precision data {\it vs.} the Higgs-boson mass, $M_{H}$.  The solid
	line is the result of the fit; the band represents an estimate of the
	theoretical error due to missing higher order corrections.  The
	vertical band shows the 95\% CL exclusion limit on $M_{H}$ from the
	direct search at LEP. The dashed curve shows the sensitivity to a
	change in the evaluation of $\alpha(M_{Z}^{2})$.  (From the LEP
	Electroweak Working Group, Ref.\ {\protect\cite{ewwg}}.)}
	\protect\label{fig:LEPhiggsfit}
\end{figure}
Also indicated in Figure \ref{fig:LEPmwmt} are the direct 
determinations of $m_{t}$ and $M_{W}$, and the values inferred from a 
universe of electroweak observables, both shown as 
one-standard-deviation regions.  The direct and indirect 
determinations are in reasonable agreement.  Both favor a light Higgs 
boson, \textit{within the framework of the standard-model analysis.}

The $M_{W}$-$m_{t}$-$M_{H}$ correlation will be telling over the next 
few years, since we anticipate that measurements at the Tevatron and 
LHC will determine $m_{t}$ within 1~or~$2\gevcc$ and improve the 
uncertainty on $M_{W}$ to about $15\mevcc$.  As the Tevatron's 
integrated luminosity grows past $10\fb^{-1}$, CDF and D\O\ will begin 
to explore the region of Higgs-boson masses not excluded by the LEP 
searches.\cite{Carena:2000yx}  Soon after that, ATLAS and CMS will 
carry on the exploration of the Higgs sector at the Large Hadron 
Collider.

By itself, the $W$-boson mass suggests a preference---always within 
the standard model---for a light Higgs boson.  The indication becomes 
somewhat stronger when all the electroweak observables are examined.  
Figure \ref{fig:LEPhiggsfit} shows how the goodness of the 
LEP Electroweak Working Group's 
global fit depends upon $M_{H}$.  Within the standard model, they 
deduce a 95\% CL upper limit, $M_{H} \ltap 170\gevcc$.  Since the direct 
searches at LEP have concluded that $M_{H}> 113.5\gevcc$, excluding 
much of the favored region, either the Higgs boson is just around the 
corner, or the standard-model analysis is misleading.  
Things will soon be popping!

\section{The electroweak scale and beyond}
We have seen that the scale of electroweak symmetry breaking, $v =
(G_{F}\sqrt{2})^{-\frac{1}{2}} \approx 246\gev$, sets the values of
the $W$- and $Z$-boson masses.  But the electroweak scale is not the
only scale of physical interest.  It seems certain that we must also
consider the Planck scale, derived from the strength of Newton's
constant, and it is also probable that we must take account of the
$SU(3)_{c}\otimes SU(2)_{L}\otimes U(1)_{Y}$ unification scale around
$10^{15\mathrm{-}16}\gev$.  There may well be a distinct flavor scale. 
The existence of other significant energy scales is behind the famous
problem of the Higgs scalar mass: how to keep the distant scales from
mixing in the face of quantum corrections, or how to stabilize the
mass of the Higgs boson on the electroweak scale.

\subsection{Why is the electroweak scale small?}
To this point, we have outlined the electroweak theory, 
emphasized that the need for a Higgs boson (or substitute) is quite 
general, and reviewed the properties of the standard-model Higgs 
boson.  By considering a thought experiment, gauge-boson scattering 
at very high energies, we found a first signal for the importance of 
the 1-TeV scale.  Now, let us explore another path to the 1-TeV scale.

The $SU(2)_L \otimes U(1)_Y$ electroweak theory does not explain how the 
scale of electroweak symmetry breaking is maintained in the presence 
of quantum corrections.  The problem of the scalar sector can be 
summarized neatly as follows.\cite{10}  The Higgs potential is
\begin{equation}
      V(\phi^\dagger \phi) = \mu^2(\phi^\dagger \phi) +
\abs{\lambda}(\phi^\dagger \phi)^2 \;.
\end{equation}
With $\mu^2$ chosen to be less than zero, the electroweak symmetry is 
spontaneously broken down to the $U(1)$ of electromagnetism, as the 
scalar field acquires a vacuum expectation value that is fixed by the low-energy
phenomenology, 
\begin{equation}
	\vev{\phi} = \sqrt{-\mu^2/2|\lambda|} \equiv (G_F\sqrt 8)^{-1/2}
		\approx 175 {\rm \;GeV}\;.
		\label{hvev}
\end{equation}

Beyond the classical approximation, scalar mass parameters receive 
quantum corrections from loops that contain particles of spins 
$J=1, 1/2$, and $0$:
\begin{equation}
\BoxedEPSF{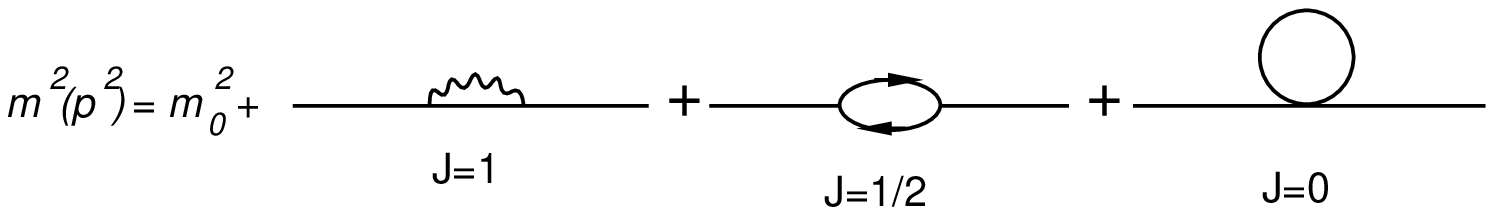  scaled 750}
\label{loup}
\end{equation}
The loop integrals are potentially divergent.  Symbolically, we may 
summarize the content of \eqn{loup} as
\begin{equation}
	m^2(p^2) = m^2(\Lambda^2) + Cg^2\int^{\Lambda^2}_{p^2}dk^2 
	+ \cdots \;,
	\label{longint}
\end{equation}
where $\Lambda$ defines a reference scale at which the value of 
$m^2$ is known, $g$ is the coupling constant of the theory, and the 
coefficient $C$ is calculable in any particular theory.  
Instead of dealing with the relationship between observables and 
parameters of the Lagrangian, we choose to describe the variation of 
an observable with the momentum scale.  In order for the mass shifts 
induced by radiative corrections to remain under control (\ie , not to 
greatly exceed the value measured on the laboratory scale), either 
$\Lambda$ must be small, so the range of integration is not 
enormous, or new physics must intervene to cut off the integral.

If the fundamental interactions are described by an 
$SU(3)_c\otimes SU(2)_L\otimes U(1)_Y$ gauge symmetry, \ie, by quantum
chromodynamics and the electroweak theory, then the 
natural reference scale is the Planck mass,\footnote{It is because
$M_{\mathrm{Planck}}$ is so large (or because $G_{\mathrm{Newton}}$ is
so small) that we normally consider gravitation irrelevant for
particle physics.  The graviton-quark-antiquark coupling is
generically $\sim E/M_{\mathrm{Planck}}$, so it is easy to make a
dimensional estimate of the branching fraction for a gravitationally
mediated rare kaon decay: $B(K_{L} \to \pi^{0}G) \sim
(M_{K}/M_{\mathrm{Planck}})^{2} \sim 10^{-38}$, which is truly
negligible!}

\begin{equation}
	\Lambda \sim M_{\rm Planck}  = 
	\left(\frac{\hbar c}{G_{\mathrm{Newton}}}\right)^{1/2} \approx 1.22 
	\times 10^{19} {\rm \; GeV}\;.
\end{equation}
In a unified theory of the strong, weak, and electromagnetic 
interactions, the natural scale is the unification scale,
\begin{equation}
	\Lambda \sim M_U \approx 10^{15}\hbox{-}10^{16} {\rm \; GeV}\;.
\end{equation}
Both estimates are very large compared to the scale of electroweak 
symmetry breaking \eqn{hvev}.  We are therefore assured that new physics must 
intervene at an energy of approximately 1~TeV, in order that the 
shifts in $m^2$ not be much larger than \eqn{hvev}.

Only a few distinct scenarios for controlling the 
contribution of the integral in \eqn{longint} can be envisaged.  The 
supersymmetric solution is especially 
elegant.\cite{lykken,SteveM,Sallyd,HMtasi}  Exploiting the fact 
that fermion loops contribute with an overall minus sign (because of 
Fermi statistics), supersymmetry balances the contributions of fermion 
and boson loops.  In the limit of unbroken supersymmetry, in which the 
masses of bosons are degenerate with those of their fermion 
counterparts, the cancellation is exact:
\begin{equation}
	\sum_{{i={\rm fermions \atop + bosons}}}C_i\int dk^2 = 0\;.
\end{equation}
If the supersymmetry is broken (as it must be in our world), the 
contribution of the integrals may still be acceptably small if the 
fermion-boson mass splittings $\Delta M$ are not too large.  The 
condition that $g^2\Delta M^2$ be ``small enough'' leads to the 
requirement that superpartner masses be less than about 
$1\tevcc$.

A second solution to the problem of the enormous range of integration in 
\eqn{longint} is offered by theories of dynamical symmetry breaking such as 
technicolor.\cite{Chivukula:1996uy,Chivukula:2000mb,etc} In technicolor models, the Higgs boson is composite, and 
new physics arises on the scale of its binding, $\Lambda_{\mathrm{TC}} \simeq 
O(1~{\rm TeV})$. Thus the effective range of integration is cut off, and 
mass shifts are under control.

A third possibility is that the gauge sector becomes strongly 
interacting. This would give rise to $WW$ resonances, multiple 
production of gauge bosons, and other new phenomena at energies of 1 TeV 
or so.  It is likely that a scalar bound state---a quasi-Higgs 
boson---would emerge with a mass less than about 
$1\tevcc$.\cite{Chanowitz:1998wi}

We cannot avoid the conclusion that some new physics must occur on 
the \onetev.\footnote{Since the superconducting phase transition 
informs our understanding of the Higgs mechanism for electroweak
symmetry breaking, it may be useful to look to other collective
phenomena for inspiration.  Although the implied gauge-boson masses
are unrealistically small, chiral symmetry breaking in QCD can induce
a dynamical breaking of electroweak symmetry.{\protect \cite{marvin}}
(This is the prototype for technicolor models.)  Is it possible that
other interesting phases of QCD---color superconductivity,{\protect
\cite{colorsc,fwhd,Rajagopal:2000wf}} for example---might hold lessons
for electroweak symmetry breaking under normal or unusual conditions? 
}

\subsection{Why is the Planck scale so large?}
\begin{figure}[tb] 
\centerline{\BoxedEPSF{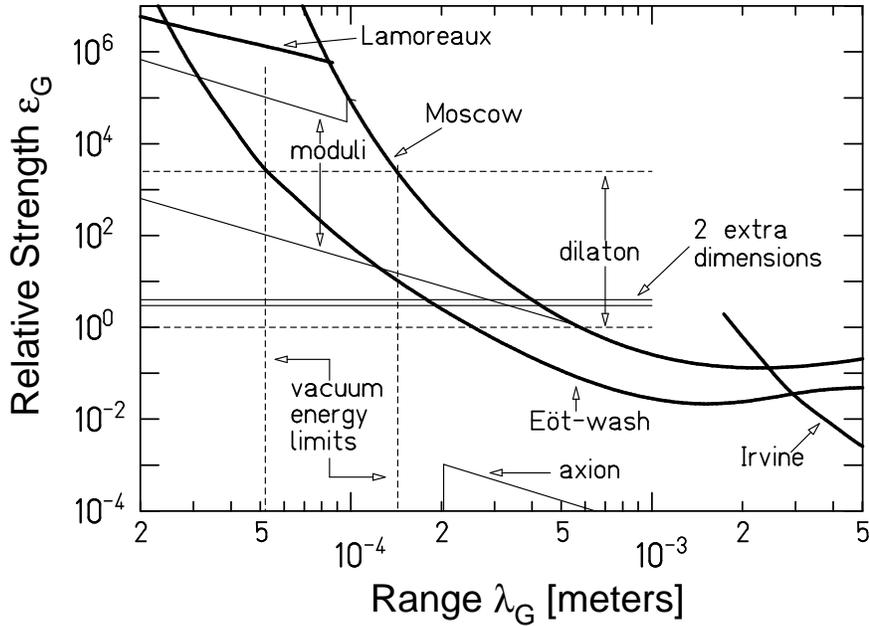 scaled 800}}
\vspace{10pt}
\caption{95\% confidence level upper limits (heavy line labeled
E\"ot-wash) on the strength $\varepsilon_{\mathrm{G}}$ (relative to
gravity) versus the range $\lambda_{\mathrm{G}}$ of a new long-range
force characterized by {\protect{\eqn{eq:nonNewt}}}, from Ref.\
{\protect\cite{erica}}.  The region excluded by earlier
work\protect\cite{ho:85,mi:88,la:98} lies above the heavy lines
labeled Irvine, Moscow and Lamoreaux, respectively.  The theoretical
predictions are adapted from Ref.~\protect\cite{pr:99}, except for the
dilaton prediction, from Ref.~\protect\cite{ka:00}.
}
\label{fig:nonNgrav}
\end{figure}

The conventional approach to new physics has been to extend the 
standard model to understand why the electroweak scale (and the mass 
of the Higgs boson) is so much smaller than the Planck scale.  A novel 
approach that has been developed over the past two years is instead to 
\textit{change gravity} to understand why the Planck scale is so much 
greater than the electroweak scale.\cite{EDbiblio}  Now, experiment 
tells us that gravitation closely follows the Newtonian force law down 
to distances on the order of $1\mm$.  Let us parameterize deviations 
from a $1/r$ gravitational potential in terms of a relative strength 
$\varepsilon_{\mathrm{G}}$ and a range $\lambda_{\mathrm{G}}$, so that
\begin{equation}
V(r) = - \int dr_{1}\int dr_{2} 
\frac{G_{\mathrm{Newton}}\rho(r_{1})\rho(r_{2})}{r_{12}} \left[ 1+ 
\varepsilon_{\mathrm{G}}\exp(-r_{12}/\lambda_{\mathrm{G}}) \right]\; ,
\label{eq:nonNewt}
\end{equation}
where $\rho(r_{i})$ is the mass density of object $i$ and $r_{12}$ is 
the separation between bodies 1 and 2.
Elegant experiments that study details of Casimir and Van der Waals 
forces imply bounds on anomalous gravitational interactions, as shown 
in Figure \ref{fig:nonNgrav}.  Below about a millimeter, the 
constraints on deviations from Newton's inverse-square force law deteriorate 
rapidly, so nothing prevents us from considering changes to gravity 
even on a small but macroscopic scale.

%%%%%%%%%%%%%%%%%%%%%%%%%%%%%%%%%%%%%%%%%%%%%%%%%%%%%%%%%%%%%%%
%                                                             %
%   \begin{equation}                                          %
%   V(r)=-G \frac {m_1 m_2}{r}(1 + \alpha e^{-r/\lambda})~.   %
%   \label{eq: potential}                                     %
%   \end{equation}                                            %
%                                                             %
%%%%%%%%%%%%%%%%%%%%%%%%%%%%%%%%%%%%%%%%%%%%%%%%%%%%%%%%%%%%%%%

For its internal consistency, string theory requires an additional six 
or seven space dimensions, beyond the $3+1$ dimensions of everyday 
experience.\cite{JLt2K}  Until recently it has been presumed that the extra 
dimensions must be compactified on the Planck scale, with a
compactification radius $R_{\mathrm{unobserved}} \approx
1/M_{\mathrm{Planck}} \approx 1.6 \times 10^{-35}\m$.  The new wrinkle 
is to consider that the $SU(3)_{c}\otimes SU(2)_{L}\otimes U(1)_{Y}$
standard-model gauge fields, plus needed extensions, reside on 
$3+1$-dimensional branes, not in the extra dimensions, but that 
gravity can propagate into the extra dimensions.

\begin{figure}[tb] 
\centerline{\BoxedEPSF{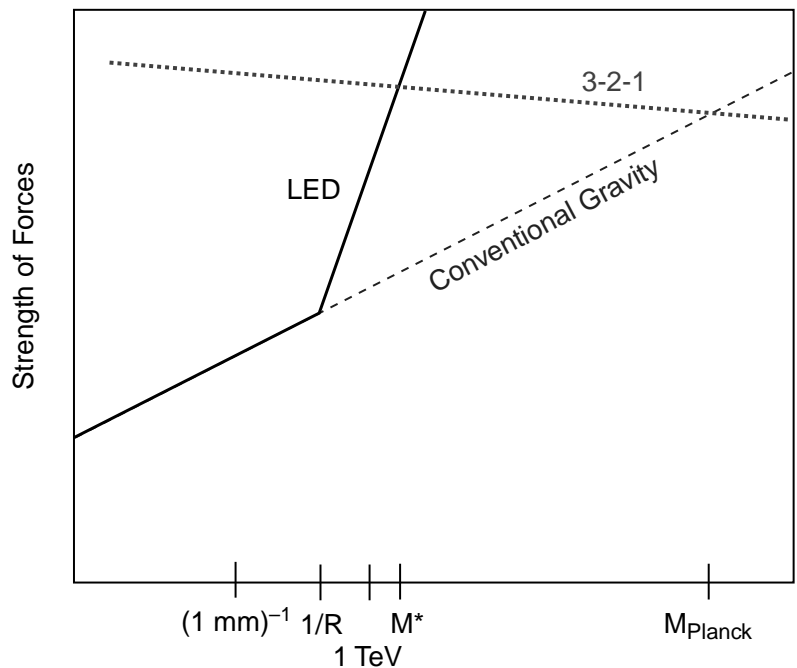 scaled 750}}
\vspace{10pt}
\caption{One of these extrapolations (at least!) is false.}
\label{fig:false}
\end{figure}

How does this hypothesis change the picture?  The dimensional 
analysis (Gauss's law, if you like) that relates Newton's constant to 
the Planck scale changes.  If gravity propagates in $n$ extra 
dimensions with radius $R$, then
\begin{equation}
    G_{\mathrm{Newton}} \sim M_{\mathrm{Planck}}^{-2} \sim M^{\star\,-n-2}R^{-n}\; ,
    \label{eq:gauss}
\end{equation}
where $M^{\star}$ is gravity's true scale.  Notice that if we boldly 
take $M^{\star}$ to be as small as $1\tevcc$, then the radius of the extra 
dimensions is required to be smaller than about $1\mm$, for $n \ge 
2$.  If we use the four-dimensional force law to extrapolate the 
strength of gravity from low energies to high, we find that gravity 
becomes as strong as the other forces on the Planck scale, as shown 
by the dashed line in Figure \ref{fig:false}.  If the force law 
changes at an energy $1/R$, as the large-extra-dimensions scenario 
suggests, then the forces are unified at an energy $M^{\star}$, as 
shown by the solid line in Figure \ref{fig:false}.
What we know as the Planck scale is then a mirage that results 
from a false extrapolation: treating gravity as four-dimensional down 
to arbitrarily small distances, when in fact---or at least in this 
particular fiction---gravity propagates in $3+n$ spatial dimensions.  
The Planck mass is an artifact, given by $M_{\mathrm{Planck}} = 
M^{\star}(M^{\star}R)^{n/2}$. 

Although the idea that extra dimensions are just around the 
corner---either on the submillimeter scale or on the TeV scale---is 
preposterous, it is not ruled out by observations.  For that reason 
alone, we should entertain ourselves by entertaining the 
consequences.  Many authors have considered the gravitational 
excitation of a tower of Ka\l uza--Klein modes in the extra 
dimensions, which would give rise to a missing (transverse) energy 
signature in collider experiments.\cite{smaria}  We call these excitations 
\textit{provatons,} after the Greek word for a sheep in a 
flock.

The electroweak scale is nearby; indeed, it is coming within 
experimental reach at LEP2, the Tevatron Collider, and the Large 
Hadron Collider.  Where are the other scales of significance?  In 
particular, what is the energy scale on which the properties of quark 
and lepton masses and mixings are set?  The similarity between the 
top-quark mass, $m_{t} \approx 175\gevcc$, and the Higgs-field vacuum 
expectation value, $v/\sqrt{2} \approx 176\gev$, encourages the hope 
that in addition to decoding the puzzle of electroweak symmetry 
breaking in our explorations of the \onetev, we might gain insight 
into the problem of fermion mass.  This is an area to be defined over 
the next decade.
\section{Outlook}
The creation of the electroweak theory is one of the great 
achievements of twentieth-century science.  Its history shows the 
importance of the interplay between theory and experiment, and the 
significance of both search-and-discovery experiments and precision, 
programmatic measurements.  We owe a special salute to the heroic 
achievements of the LEP experiments, of the LEP accelerator team, and 
of the theorists who devoted themselves to extracting the most (reliable!) 
information from precision measurements.  Their work has been 
inspiring.

The electroweak story is not done.  We haven't fully understood the 
agent of electroweak symmetry breaking, but we look forward to a 
decade of discovery on the \onetev.  We probably have much to learn 
about the nonperturbative aspects of the electroweak theory, about 
which I've said nothing in these lectures.\cite{riotr}  And we are 
just learning to define---then to solve---the essential mystery of 
flavor, the problem of identity.  These are the good old days!

For many topics not covered in these lectures, including an
introduction to Higgs searches, see my 1998 Granada lectures.\cite{Quigg:1999xg}

\section*{Acknowledgments}
Fermilab is operated by Universities Research Association Inc.\ under
Contract No.\ DE-AC02-76CH03000 with the United States Department of
Energy.  It is a pleasure to thank Steve Ellis for gracious 
hospitality at the University of Washington during the writing of 
these notes.  I would also like to take this opportunity to 
compliment Jonathan Rosner, Hitoshi Murayama, K. T. Mahanthappa, and 
the TASI staff for their efficient organization of ``Flavor Physics for 
the Millennium,'' and for the pleasant and stimulating atmosphere in Boulder.


\begin{thebibliography}{99}
\bibitem{EQT} In the high-energy limit, an amplitude for longitudinal 
gauge-boson interactions may be replaced by a corresponding amplitude 
for the scattering of massless Goldstone bosons: $\mathcal{M}(W_{L},Z_{L}) = 
\mathcal{M}(w,z) + \mathcal{O}(M_{W}/\sqrt{s})$ The 
equivalence theorem can be traced to the work of John M.\  Cornwall, 
David N.\  Levin, and George Tiktopoulos, \pr{10}{1145}{74}, 
\ib{11}{972E}{75}.  It was applied to this problem by Lee, Quigg, and 
Thacker, Ref.\ \cite{lqt}, and developed extensively by Chanowitz and 
Gaillard, Ref. \cite{LT8}, and others.
\end{thebibliography}
\end{document}